\documentclass[]{aa}

\usepackage[dvips]{graphicx}
\usepackage{amsmath}
\usepackage{amssymb}

\begin{document}

\title{A multiwavelength study of the massive star forming region IRAS
06055+2039 (RAFGL 5179)}
\author{A. Tej\inst{1}, D. K. Ojha\inst{1}, S. K. Ghosh\inst{1}, V. K.
Kulkarni\inst{2}, R. P. Verma\inst{1}, S. Vig\inst{1} and T. P.
Prabhu\inst{3}}

\offprints{A. Tej, email: tej@tifr.res.in}

\institute{
Tata Institute of Fundamental Research, Mumbai 400 005, India\\
\and
National Centre for Radio Astrophysics, Post Bag 3, Ganeshkhind, Pune
411 007, India\\
\and
Indian Institute of Astrophysics, Koramangala, Bangalore 560 034,
India
}
\date{Received xxx / Accepted yyy}

\authorrunning{A. Tej et al.}
\titlerunning{Multiwavelength study of IRAS 06055+2039}
\abstract{
{We present a multiwavelength study of the massive star forming region
associated with IRAS 06055+2039 which reveals an interesting scenario
of this complex where regions are at different stages of evolution of
star formation. 
Narrow band near-infrared (NIR)
observations were carried out with UKIRT-UFTI in molecular hydrogen and Br$\gamma$
lines to trace the shocked and ionized gases respectively. 
We have used 2MASS $J H K_{s}$ data to study the nature of the
embedded cluster associated with IRAS 06055+2039. We obtain a
power-law slope of 0.43$\pm$0.09 for the $K_{s}$-band Luminosity Function
(KLF) which is in good agreement with other young embedded clusters. 
We estimate an age of 2 -- 3 Myr for this cluster.
The radio emission from the ionized gas has been mapped at 610 and 1280 MHz
using the Giant Metrewave Radio Telescope (GMRT), India. 
Apart from the diffuse emission, the high resolution 1280 MHz
map also shows the presence of several discrete sources which possibly
represent high density clumps. 
The morphology of shocked molecular hydrogen forms an arc towards the N-E
of the central IRAS point source and envelopes the radio emission.
Submillimetre emission using JCMT-SCUBA show the presence of a
dense cloud core which is probably at an earlier evolutionary stage
compared to the ionized region with shocked molecular gas lying in
between the two.
Emission from warm dust and the Unidentified Infrared Bands (UIBs) have been estimated
using the mid-infrared (8 -- 21\,$\mu$m) data from the MSX survey.
From the submillimetre emission at 450 and 850\,$\mu$m 
the total mass of the cloud is estimated
to be $\sim$ 7000 -- 9000 $\rm M_{\odot}$. 
}
}

\maketitle

\section{Introduction}
Massive stars are
preferentially formed in dense cores of molecular clouds.
They remain deeply embedded in the prenatal molecular gas and
obscuring dust and their pre-main sequence time scales are much shorter
compared to the low mass stars. The luminous high mass stars also affect 
the parent cloud. In addition, massive
stars do not form in isolation but often in clusters and
associations. All these factors contribute in making the study of the
formation mechanisms of these systems very difficult. Multiwavelength
studies, therefore, hold the potential to probe these complexes at
different depths and unravel the least understood aspects of 
massive star formation.
IRAS 06055+2039 (G189.78+0.34, RAFGL 5179) is a massive star forming region chosen from the
catalog of massive young stellar objects by Chan et al. (1996).
G189.78+0.34 is listed as an ultracompact (UC) HII region
(Shepherd \& Churchwell 1996; Bronfman et al. 1996).
It belongs to the Gem OB1 molecular cloud complex and is a part of the extended HII region
Sh 252. It is associated with S252 A which is one of the six compact
radio sources in Sh 252 revealed from the 5 GHz aperture synthesis observations by
Felli et al. (1977). 
IRAS fluxes yield a far infrared luminosity of $\sim 10^{4}\rm
L_{\odot}$ (Carpenter et al. 1995) for this star forming region.
There are several kinematic distance estimates used in literature for
this source which range from 1.25 kpc (Mirabel et al. 1987) to 2.9 kpc
(Wouterloot \& Brand 1989). In this paper we use the value of 2.6 kpc
(Wu et al. 2001) which is the most widely used distance estimate
for this source. 

This high mass star forming region has been observed as part of many surveys. 
H$_{2}$O maser (K\"{o}mpe et al. 1989; Lada \& Wooden 1979)
and the 6.7 GHz methanol maser (Szymczak et al. 2000a) have been detected towards IRAS
06055+2039. Positive detection has also been made in SiO (Harju et al.
1998), CO (2 - 1) (Wu et al. 2001) and CS (2 - 1) (Bronfman et al.
1996). Zinchenko et al. (1998) in their study of dense molecular cores
have also mapped this source in CS (2 - 1). 
CO maps of Shepherd \& Churchwell (1996) do not show evidence of any
bipolar outflows. Search for the 6 cm (Szymczak et al. 2000b) and 5 cm 
(Baudry et al. 1997) OH masers show negative results. 
The IRAS low resolution spectra show a relatively red continuum from
13 to 23\,$\mu$m with the presence of an emission feature at
11.3\,$\mu$m which is attributed to the presence of Polycyclic
Aromatic Hydrocarbon (PAH) molecules (Kwok et al. 1997).
    
In this paper, we present a multiwavelength study of this star forming
region. 
In Sect.\,\,\ref{obvs.sec}, we present the narrow-band near-infrared
(NIR) and radio continuum observations and a brief description of the
related data reduction procedures. In
Sect.\,\,\ref{arch.sec}, we discuss other available datasets
used in the present study. Section\,\,\ref{results.sec} gives a
comprehensive discussion on the results obtained and in
Sect.\,\,\ref{concl.sec}, we summarize the results.

%--------------------------------------------------------------------
\section{Observations and Data Reduction}
\label{obvs.sec}

\subsection{Near-Infrared Observations}
\begin{figure*}
\centering
\resizebox{\hsize}{!}{\includegraphics{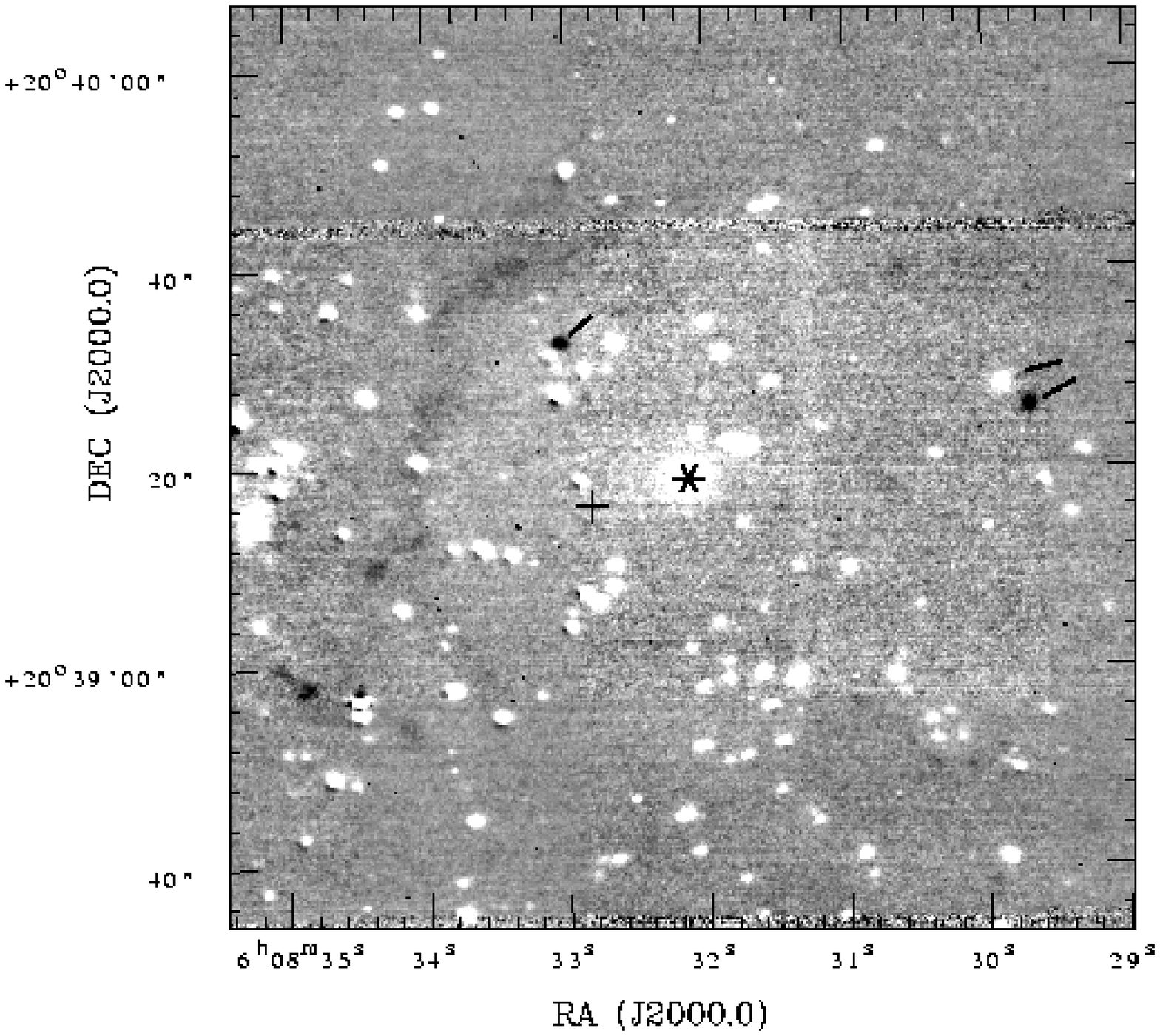}\includegraphics{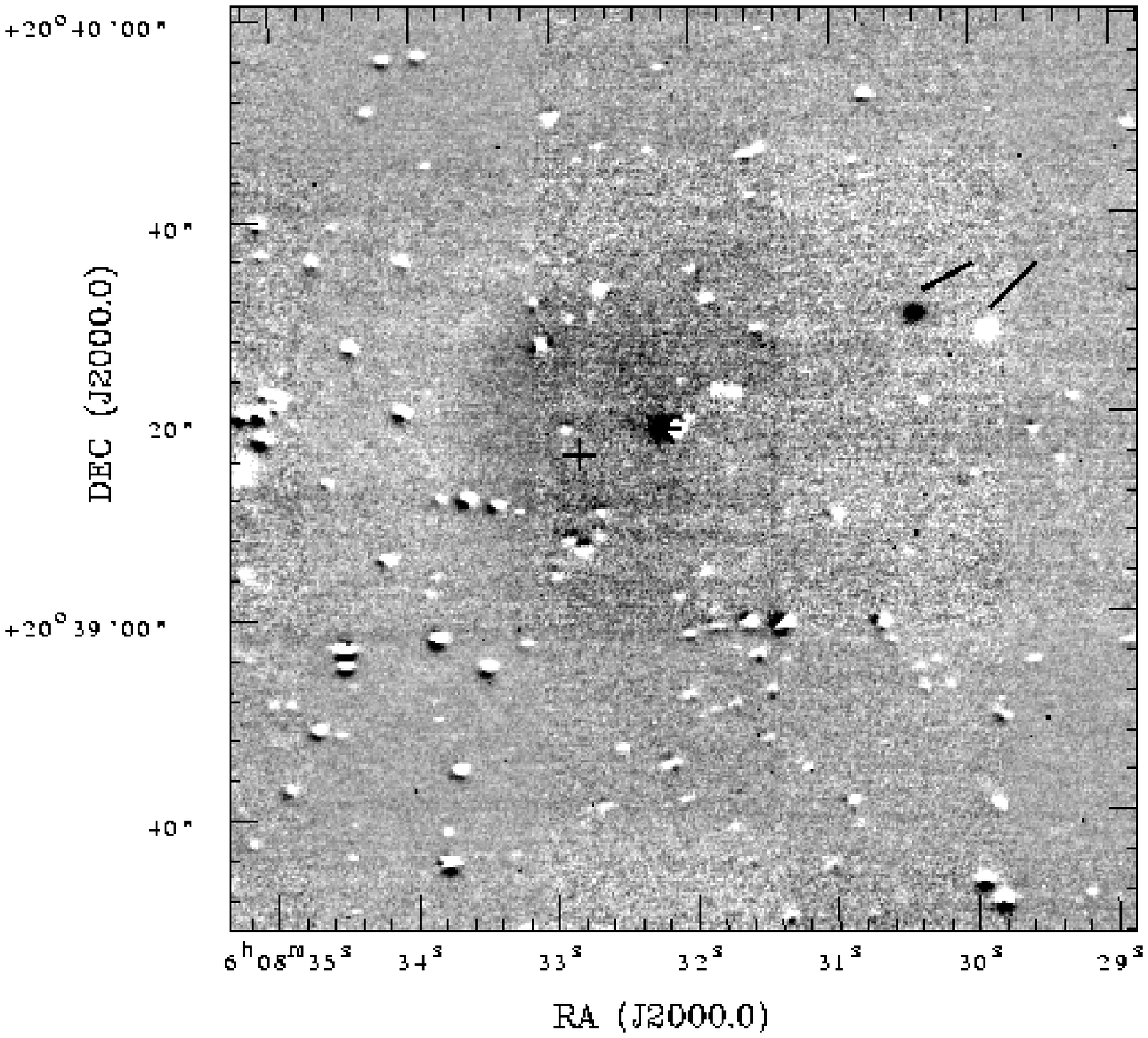}}
\caption{The continuum subtracted H$_{2}$ (1-0)S1 (left) and Br$\gamma$
(right) images of the field around IRAS 06055+2039. The images were
smoothed to a resolution of $\rm 0.3\arcsec \times 0.3\arcsec$ ($\rm
3 \times 3$ box) to improve the contrast of the faint
features. Residuals of continuum subtraction around stars are seen in both the
images. The plus sign and the asterisk in both the images mark the position of the
IRAS point source and the central brightest star in the NIR cluster
respectively. Artifacts (possibly ghost images of the central bright star)
are seen in H$_{2}$, Br$\gamma$ and the $K$ continuum frames. The
resulting residuals in both the continuum subtracted images are indicated with
lines.}

\label{h2_brg.fig}
\end{figure*}

Narrow-band NIR observations were carried out in the
rotational-vibrational line of molecular hydrogen -- H$_{2}$ (1-0)S1 ($\lambda$=2.12\,$\mu$m,
$\Delta\lambda$=0.02\,$\mu$m), hydrogen recombination line of Br$\gamma$
($\lambda$=2.16\,$\mu$m, $\Delta\lambda$=0.02\,$\mu$m) and $K$ continuum
($\lambda$=2.27\,$\mu$m, $\Delta\lambda$=0.02\,$\mu$m) at the
3.8m United Kingdom Infrared Telescope (UKIRT\footnote{The United Kingdom 
Infrared Telescope is operated by the Joint Astronomy Centre on behalf of 
the U.K. Particle Physics and Astronomy Research Council.}), Hawaii. 
Observations were carried out on 16 Feb 2002 under the UKIRT Service 
Programme (Proposal No. 1459). The instrument used was the UKIRT Fast Track Imager
(UFTI), which is a 1 -- 2.5\,$\mu$m camera with a 1024$\times$1024 HgCdTe array having a plate scale of 
$\rm 0.091\arcsec$ per pixel. The `JITTER-SELF-FLAT' Data
Reduction (DR) recipe was used. This script takes imaging
observations comprising of 9 jittered object frames and a dark frame.
Flat field is then created from the sequence of jittered object
frames. The final image is a mosaic generated from the 9 frames after dark
subtraction and flat fielding and has a total field of view of $\sim$ 4$\times$4 arcmin$^{2}$. 
For our observations, the integration time was 100s in each band.
To obtain pure
emission line images it is essential to subtract out the contribution
from the continuum. This is done by subtracting the $K$ continuum
image from the H$_{2}$ and the Br$\gamma$ images after proper
alignment and PSF matching. 
Figure\,\,\ref{h2_brg.fig} shows the continuum subtracted H$_{2}$ and 
Br$\gamma$ images of the central field around IRAS 06055+2039. The continuum subtracted
H$_{2}$ image displays a prominent arc towards the N-E of the
central source. The continuum subtracted Br$\gamma$ image shows the
presence of faint diffuse emission surrounding the central bright
source.

%-------------------------------------------------------------------------------------------
\subsection{Radio Continuum Observations}
In order to probe the ionized gas component,
radio continuum interferometric mapping of the region around IRAS
06055+2039 was carried out using the Giant Metrewave Radio Telescope
(GMRT) array, India. The GMRT has a ``Y" shaped hybrid configuration
of 30 antennas, each of 45 m diameter. There are six antennas along
each of the three arms (with arm length of $\sim$ 14 km). These
provide high angular resolution (longest baseline $\sim$ 25 km). The
rest of the twelve antennas are located in a
random and compact $\rm 1\times1$ $\rm km^{2}$ arrangement near the
centre and is sensitive to large scale diffuse emission (shortest baseline
$\sim$ 100 m).
Details of the GMRT antennae and their configurations can be found in
Swarup et al. (1991).
Observations were carried out at 1280 and 610 MHz. 
The radio sources
3C48 and 3C147 were used as the primary flux calibrators 
while 0532+194 and 0432+416 were used as phase calibrators for the 1280
and 610 MHz observations respectively.

Data reduction was done using AIPS. The data sets were
carefully checked using tasks UVPLT and VPLOT for bad data (owing to
dead antennae, bad baselines, interference, spikes etc). Subsequent editing 
was carried out using the tasks UVFLG and TVFLG. Maps of the field were
generated by Fourier inversion and cleaning using the task IMAGR.
Several iterations of self calibration were carried out to 
obtain improved maps.

\begin{figure*}
\centering
\resizebox{\hsize}{!}{\includegraphics{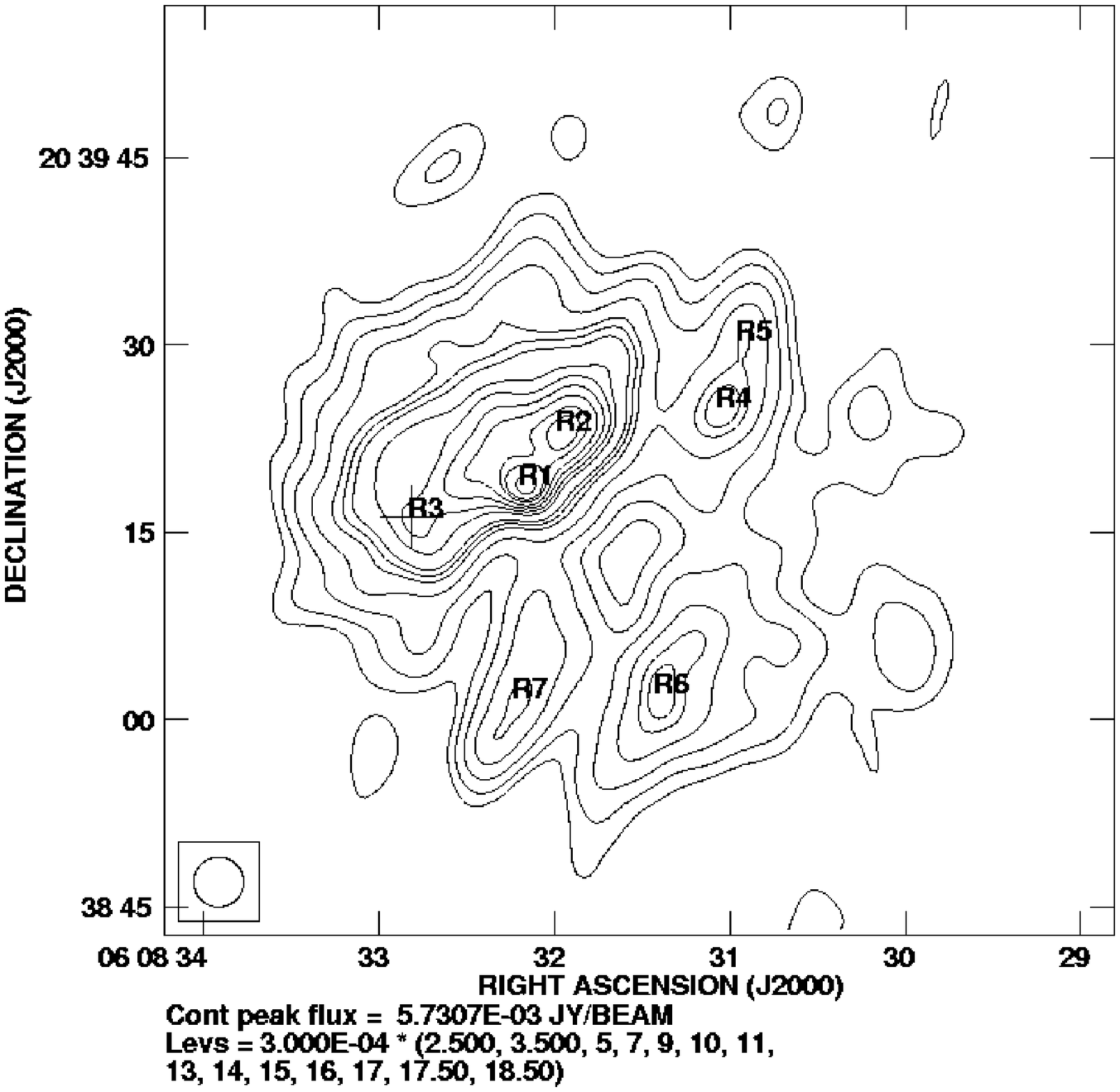}\includegraphics{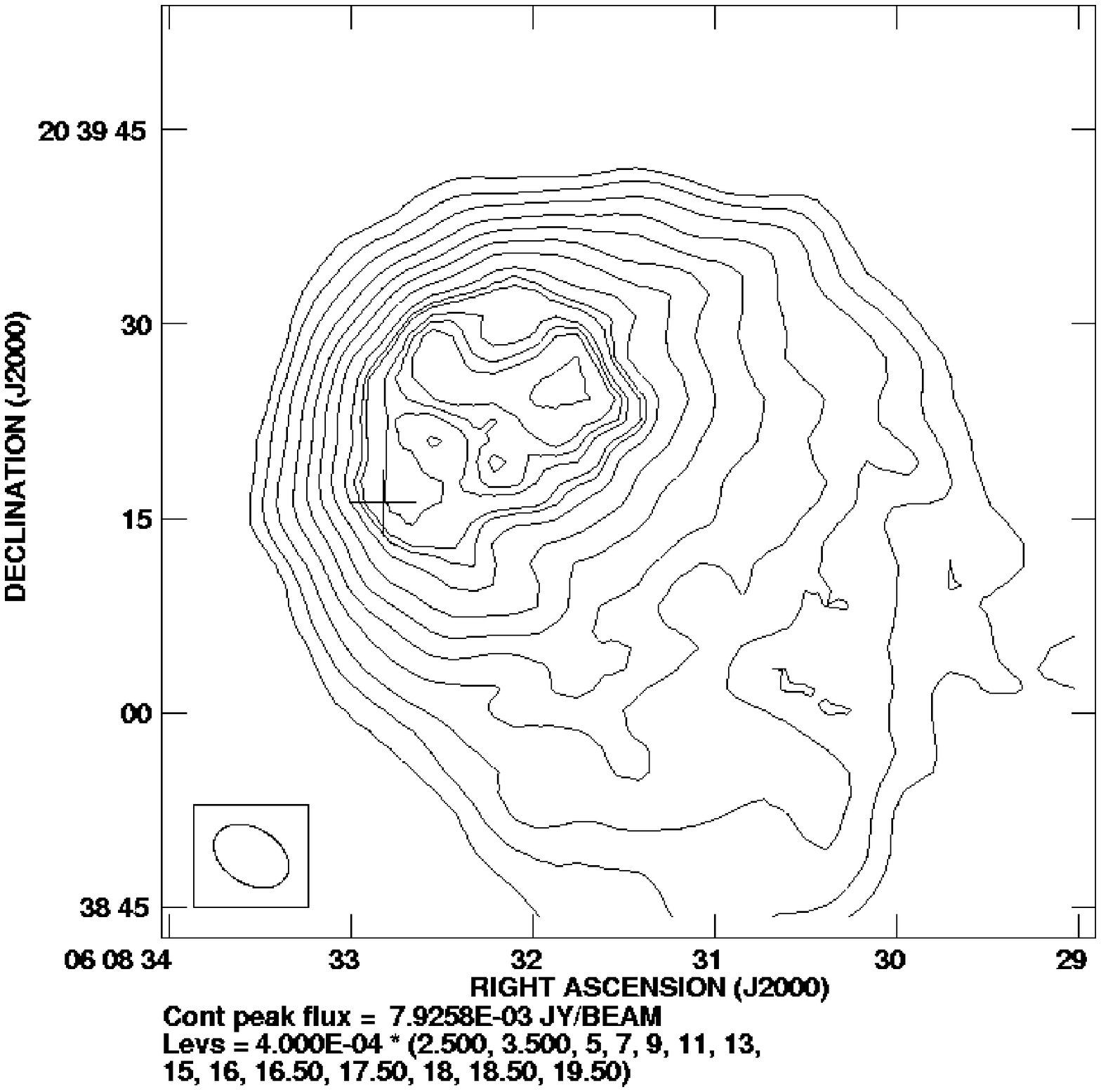}}
\caption{High resolution radio continuum maps at 1280 MHz (left) and 610 MHz
(right) for the region around IRAS 06055+2039. The
numbers R1, R2, .....R7 mark the position of discrete points sources
(see text). The rms noises at 1280 and 610 MHz are $\sim$ 0.3 and 0.4 mJy/beam
respectively. The plus sign in both the images marks the IRAS
point source position.} 
\label{1280_610.fig}
\end{figure*}

Figure\,\,\ref{1280_610.fig} shows the radio continuum images at
1280 and 610 MHz with synthesized beam sizes of
$\rm 4\arcsec.0 \times 4\arcsec.0$ and $\rm
6\arcsec.3\times4\arcsec.3$ respectively. The rms
noises in the maps are 0.3 (1280 MHz) and 0.4 (610 MHz) mJy/beam. 
Table\,\,\ref{radio.tab} gives the details of the observations and the maps. 

\begin{table}
\caption{Details of the radio interferometric continuum observations.}
\label{radio.tab}
\begin{tabular}{l|c|c}
\hline\hline
  Details         & 1280 MHz & 610 MHz \\
\hline
Date of Obs.      & 3 Aug 2003 & 21 May 2002 \\
Primary beam      & 26\arcmin & 54\arcmin \\
Cont. bandwidth (MHz) & 16        & 16  \\ 
Synth. beam   &  4\arcsec.0 $\times$ 4\arcsec.0 &
6\arcsec.3$\times$4\arcsec.3 \\
Position angle. (deg)  &  --            & 58 \\
Peak Flux (mJy/beam)& 5.7 & 7.9  \\
rms noise (mJy/beam) & 0.3            & 0.4 \\
\hline
\end{tabular}
\end{table}

%-----------------------------------------------------------------------------------------------------
\section{Other Available Datasets}
\label{arch.sec}

\subsection{Near-Infrared Data from 2MASS}
NIR ($JHK_{s}$) data for point sources around IRAS
06055+2039 have been obtained from the Two Micron All Sky
Survey\footnote{This publication makes use of data products from the Two Micron All Sky
Survey, which is a joint project of the University of Massachusetts and the
Infrared Processing and Analysis Center/California Institute of Technology,
funded by the NASA and the NSF.}
(2MASS) Point Source Catalog (PSC). Source selection was based on the `read-flag' which gives the
uncertainties in the magnitudes. In our sample we retain only those
sources for which the `read-flag' values are 1 -- 6. The 2MASS data
have been used to study
the embedded cluster associated with IRAS 06055+2039. 

\subsection{Mid-Infrared Data from MSX}
The Midcourse Space Experiment\footnote{This research made use of data
products from the Midcourse
Space Experiment. Processing of the data was funded by the Ballistic
Missile Defense Organization with additional support from NASA Office of Space
Science. This research has also made use of the NASA/ IPAC Infrared Science
Archive, which is operated by the Jet Propulsion Laboratory, Caltech, under
contract with the NASA.} (MSX) surveyed the Galactic plane
in four mid infrared bands -- A (centre $\lambda$: 8.28\,$\mu$m;
50\% peak intensity range: 6.8 -- 10.8\,$\mu$m), C (12.13$\mu$m; 11.1 -- 13.2$\mu$m),
D (14.65$\mu$m; 13.5 -- 15.9$\mu$m) and E (21.34$\mu$m; 18.2 -- 25.1$\mu$m) 
at a spatial resolution of $\rm \sim 18\arcsec$ (Price et al. 2001). 
Two of these bands (A \& C) cover the Unidentified Infrared Bands (UIBs) at 
6.2, 7.7, 8.7, 11.3 and 12.7\,$\mu$m. The integrated
flux densities of IRAS 06055+2039 in these bands are listed in Table\,\,\ref{msx_hires.tab}.
The MSX images in these four bands for the region around IRAS 06055+2039 have
been used to study the emission from the UIBs and to estimate the
spatial distribution of temperature and optical depth of the warm
interstellar dust.
\begin{table*}
\caption{Infrared flux densities for IRAS 06055+2039}
\begin{tabular}{c|c c c c | c| c c c c} 
\hline\hline 
& \multicolumn{4}{c|}{MSX$^a$ images} & \multicolumn{5}{c}{IRAS images}\\
\hline
Wavelength ($\mu$m) & 8.3 & 12.1& 14.7 & 21.3  && 12 & 25 & 60 & 100 \\ 
\hline
Flux density (Jy) & 26  & 34  & 24   & 68    &PSC & 15.6 & 76.8 & 1032 & 1715\\
                    &     &      &      &      &HIRES$^a$ & 38   &  113 & 1280 & 1691\\
\hline
\end{tabular}\\
$^a$ Flux densities were obtained by integrating over a circular
region of diameter $\rm 3\arcmin$ centered on the peak. \\
\label{msx_hires.tab}
\end{table*}

\subsection{Mid- and Far-infrared Data from IRAS}
The data from the IRAS survey in the four bands (12, 25, 60 and 100\,$\mu$m) for the
region around IRAS 06055+2039 were HIRES processed (Aumann et al. 1990) to obtain high
angular resolution maps. 
These maps have been used to determine the spatial
distribution of dust colour temperature and optical depth.
The integrated flux densities from the HIRES processed
images and IRAS-PSC are also given in Table\,\,\ref{msx_hires.tab}.

\subsection{Sub-mm Data from JCMT}
\begin{figure*}
\centering
\resizebox{\hsize}{!}{\includegraphics{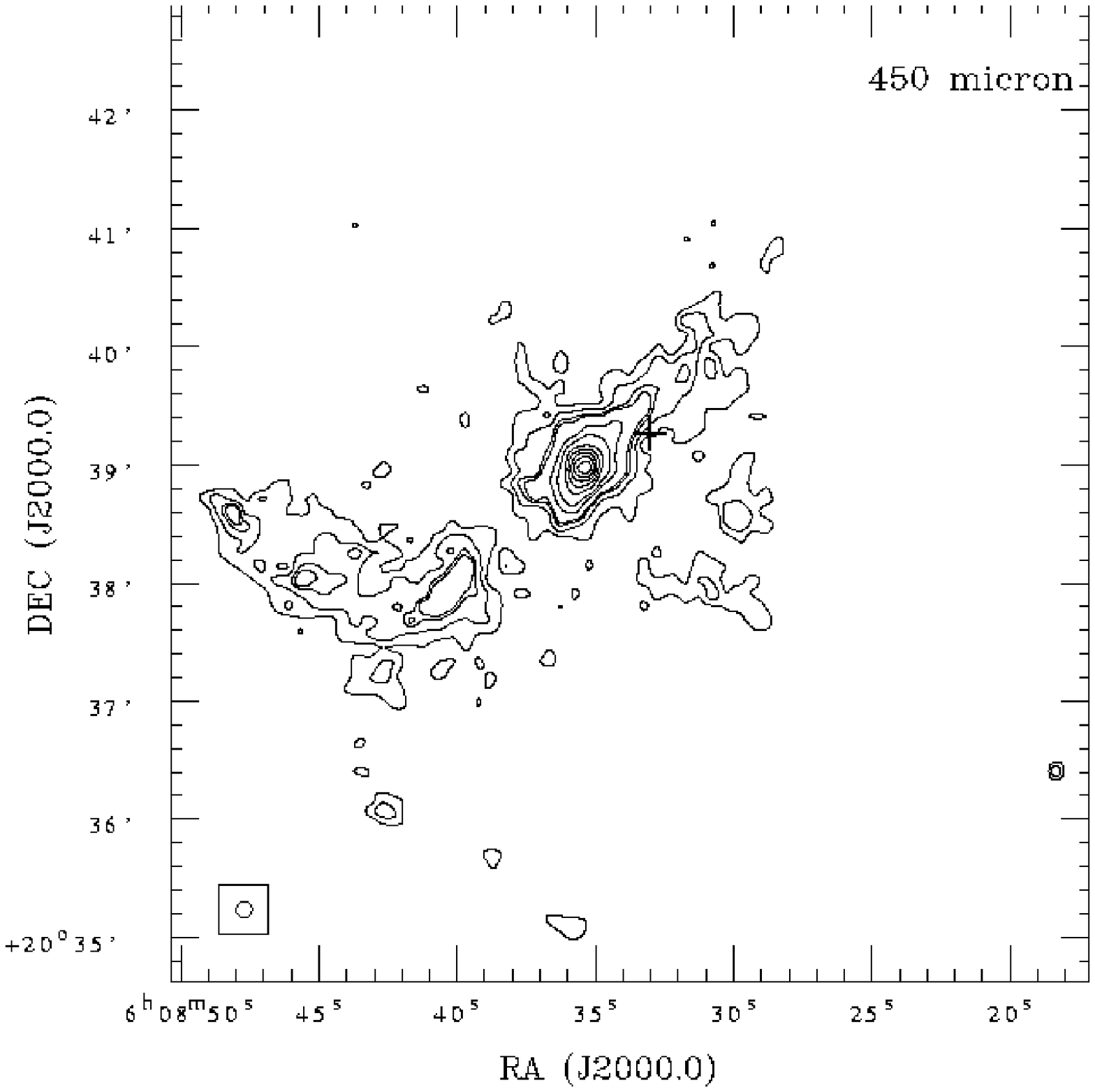} \includegraphics{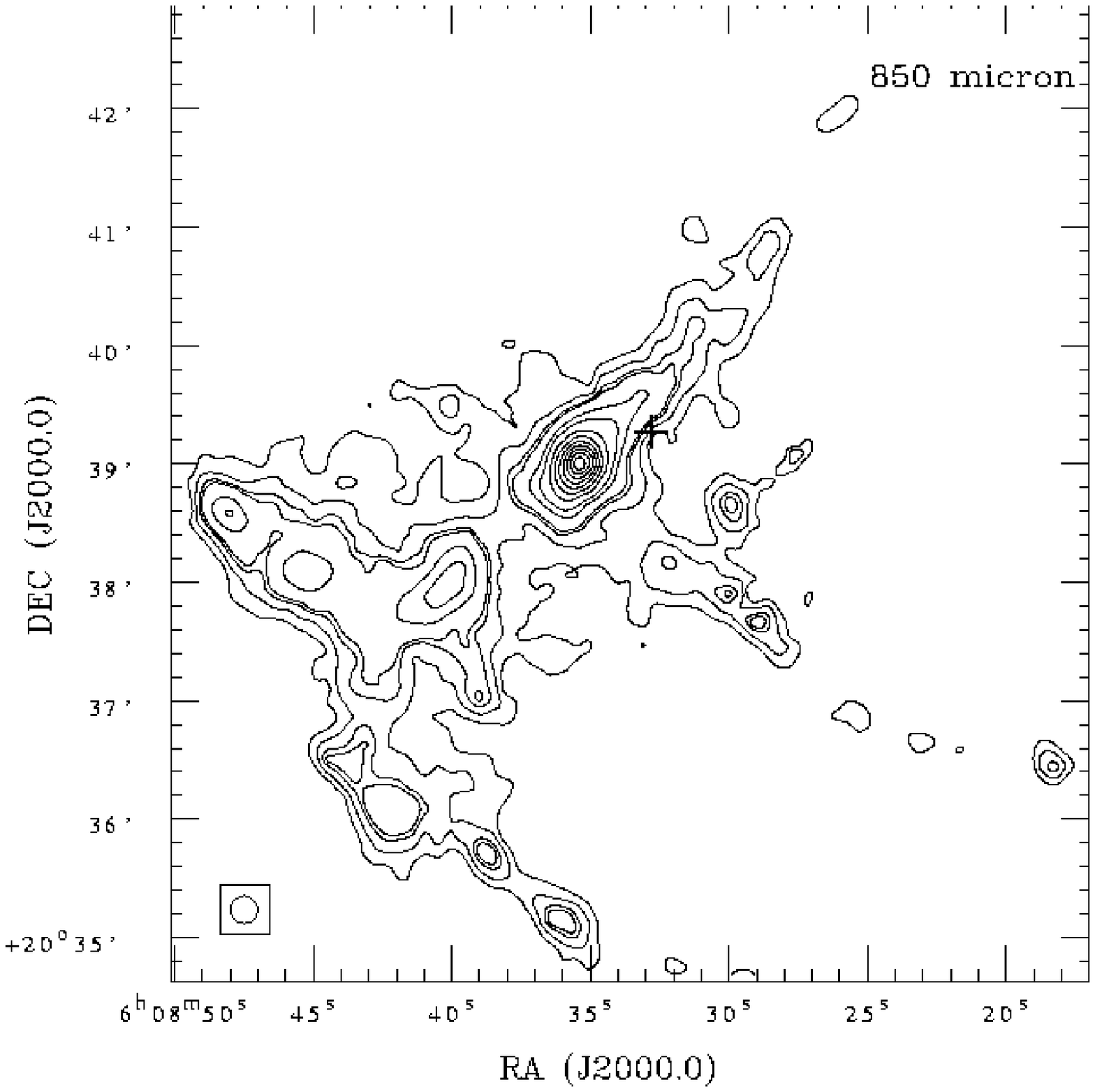}}
\caption{Contour maps showing the spatial distribution of dust emission at 450\,$\mu$m (left) and
850\,$\mu$m (right) for the region around IRAS 06055+2039.
The contour levels are at 5, 7, 9, 10, 15, 20, 30, 40, 50, 
60, 70, 80 and 90\% of the peak value of 29.45 and 6.83 Jy/beam
at 450 and 850\,$\mu$m respectively. The FWHMs of the symmetric 2-D Gaussian beams are
$\rm 7\arcsec.8$ and $\rm 15\arcsec.2$ for the two wave bands. The
plus sign in both the images marks the position of the IRAS point
source.}
\label{jcmt.fig}
\end{figure*}

Submillimetre observations using the Submillimetre Common-User Bolometer Array
(SCUBA) instrument of the James Clerk Maxwell Telescope\footnote{This paper
makes use of data from the James Clerk Maxwell Telescope Archive. The
JCMT is operated by the Joint Astronomy Centre on behalf of the UK particle
Physics and Astronomy Research Council, the National Research Council of
Canada and the Netherlands Organisation for Pure Research.} (JCMT) were retrieved
from the JCMT archives and processed using their standard pipeline SCUBA User
Reduction Facility (SURF). JCMT-SCUBA observations for the data
used in our study were carried out on 25 Oct 2000. Uranus was used as primary flux calibrator
for the maps. Figure\,\,\ref{jcmt.fig} 
displays the spatial distribution of cold dust emission at 450 and 850\,$\mu$m.
Large atmospheric extinction correction has been applied to obtain the 450\,$\mu$m
map.

%------------------------------------------------------------------------------
\section{Results and Discussion}
\label{results.sec}
\subsection{Embedded Cluster in the Near-Infrared}
\subsubsection{Radial Profile and Stellar Surface Number Density}
\label{ssnd.sect}
\begin{figure}
\centering
\resizebox{\hsize}{!}{\includegraphics{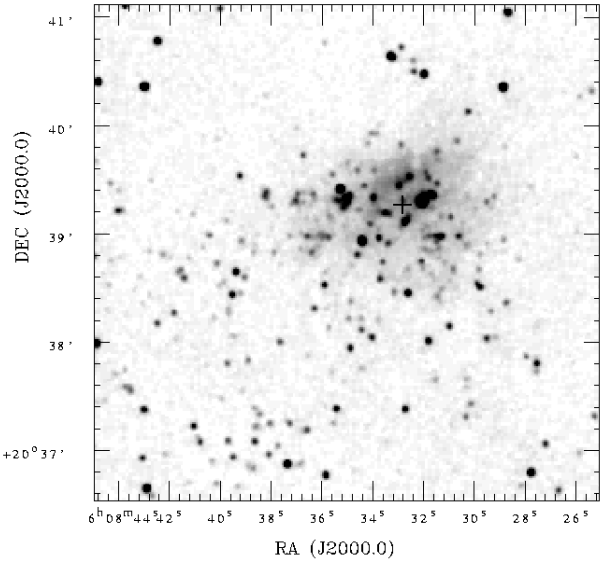}}
\caption{ The 2MASS $K_{s}$ - band image of the region around IRAS
06055+2039. The presence of a diffuse emission region and a NIR
cluster is seen. The plus sign marks the position of the IRAS point
source.}
\label{2mass_k.fig}
\end{figure}

The 2MASS $K_{s}$ - band image of the region around IRAS 06055+2039 
is shown in Figure\,\,\ref{2mass_k.fig}. 
We see the presence of a diffuse emission region
harbouring an infrared cluster. 

\begin{figure}
\resizebox{\hsize}{!}{\includegraphics{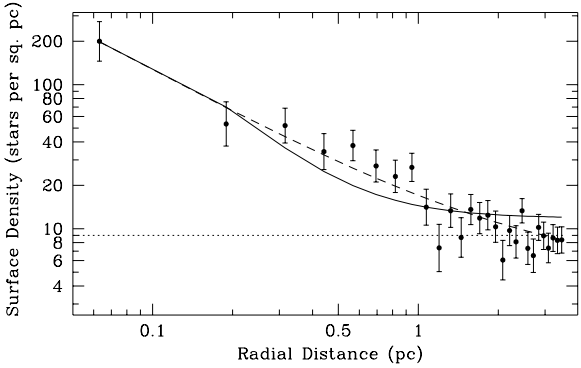}}
\caption{The radial profile of the surface number density for the cluster
associated with IRAS 06055+2039 in log-log scale. 
Also plotted are the two fitted models -- King's model (solid) and
inverse radius (dashed). As compared to the King's
model, the inverse radius model fits the radial profile much
better.
The horizontal dotted line corresponds to the background field star
level which is $\sim$ 9 stars pc$^{-2}$. Statistical errors are shown.
}
\label{cl_rad.fig}
\end{figure}

This cluster is also listed in the
catalog of embedded infrared clusters compiled by Bica et
al. (2003). We use the 2MASS $JHK_{s}$ data to study the nature of this
cluster. To determine the $K_{s}$ - band radial profile and the stellar surface
number density (SSND), we have selected sources which are detected in
the $K_{s}$ band. 
To estimate the cluster radius we select a large region of
radius $\rm 300\arcsec$ centered on IRAS 06055+2039 ($\rm
\alpha_{2000.0} = 06^{h} 08^{m}
32^{s}.1\,;\,\delta_{2000.0} = +20^{\circ} 39\arcmin
18\arcsec$).
To account for the contribution from the field stars
we select a control field ($\rm \alpha_{2000.0} = 06^{h} 09^{m}
52^{s}.0\,;\,\delta_{2000.0} = +20^{\circ} 39\arcmin
18\arcsec$) which is $\sim$
$\rm 20\arcmin$ to the east of IRAS 06055+2039.
Figure\,\,\ref{cl_rad.fig} shows the radial profile of the stellar
density in log-log scale. 
This profile was created by counting the number of stars in
$\rm 10\arcsec$ annuli and normalizing by the annulus area.
We fit two models to the surface density radial profile -- the King's
profile and the inverse radius ($r^{-1}$) model. 
Neglecting the tidal radius, the King's profile can be written as
\begin{equation}
f(r) = a + \frac{f_{0}}{1+(r/r_{c})^{2}}
\end{equation}
where, $f_{0}$ is the core concentration at radius zero, $r_{c}$ is
the core radius and $a$ is a constant for the background offset. As
seen in Fig.\,\,\ref{cl_rad.fig}, both the models describe the
density distribution fairly well. 
However, the $r^{-1}$ model has a better overall fit (reduced
$\chi^{2}$ = 1.2) as compared to the King's model (reduced $\chi^{2}$
= 3.3).
Several studies have shown that young embedded clusters have been fitted by
$r^{-1}$ profiles (e.g McCaughrean \& Stauffer 1994; Lada \& Lada
1995) or, by both the inverse radius model and the King's model (e.g
Horner et al. 1997; Teixeira et al. 2004; Baba et al. 2004). As
discussed in Baba et al. (2004), the $r^{-1}$ dependence is likely to
be a reminiscent of the parental cloud core whereas the King's profile
represents systems in dynamical equilibrium. Hence, the better fitting
of the $r^{-1}$ model could possibly suggest that the cluster associated with
IRAS 06055+2039 is not yet in complete dynamical equilibrium. 

Within errors, the cluster profile merges 
with the field star level at $\rm \sim 85\arcsec$ which translates to $\sim$ 1.1 pc 
at a distance of 2.6 kpc. We take this as the cluster radius.
The background level as estimated from the control field is $\sim \rm
9\, stars\, pc^{-2}$ which is in reasonable agreement with the value of
$\rm 11.8\pm1.6\, stars\, pc^{-2}$ yielded by the King's profile fitting. The King's profile
fitting also gives a core radius $\rm r_{c} \,\sim 0.1\, pc$.
The core radius is a scale parameter and depends mainly
on cluster parameters like density, luminosity, total mass etc. 
Several studies have shown that the core radius of the cluster is
also correlated with its age.
In their study of young clusters in the LMC, Elson et al. (1989) 
show that the core radii increases between $\sim 10^{6}$ and $10^{9}$
yr and then begin to decrease again. Such trends in core radius
evolution has also
been discussed by Wilkinson et al. (2003) for the LMC clusters and
Mackey \& Gilmore (2003) for clusters in the SMC. These authors have
shown that apart from the general increasing trend, the spread in the
core radii also increases with the age of the cluster.
Teixeira et al. (2004) and Baba et al. (2004) derive core radius values of 0.05
and 0.08 pc for the clusters NGC 2316 and RCW 36 respectively. The age
estimate for both these clusters is 2 -- 3 Myr. For a comparatively
older (5 -- 10 Myr) cluster NGC 2282, Horner et al. (1997) obtain a core radius 
of 0.19 pc. The above values of core radii seem to
suggest that the cluster associated with IRAS 06055+2039 is likely to
be of the same age ($\sim$ 2 -- 3 Myr) as NGC 2316 and RCW 36.

The number of stars detected in the $K_{s}$ band within $\rm
85\arcsec$ radius of
the cluster is 114. The total background population is 34 [$\rm = \,9
(background) \times \pi \times (1.1)^{2} (area \,of\, cluster)$].
Hence, the total number of cluster
members is estimated to be 80. This yields a space density of $\sim$
14 stars\,pc$^{-3}$. However, it should be noted here that this
density is a lower limit as we are not completely sampling the stellar
population at fainter magnitudes due to the 2MASS sensitivity limits.

\begin{figure}
\resizebox{\hsize}{!}{\includegraphics{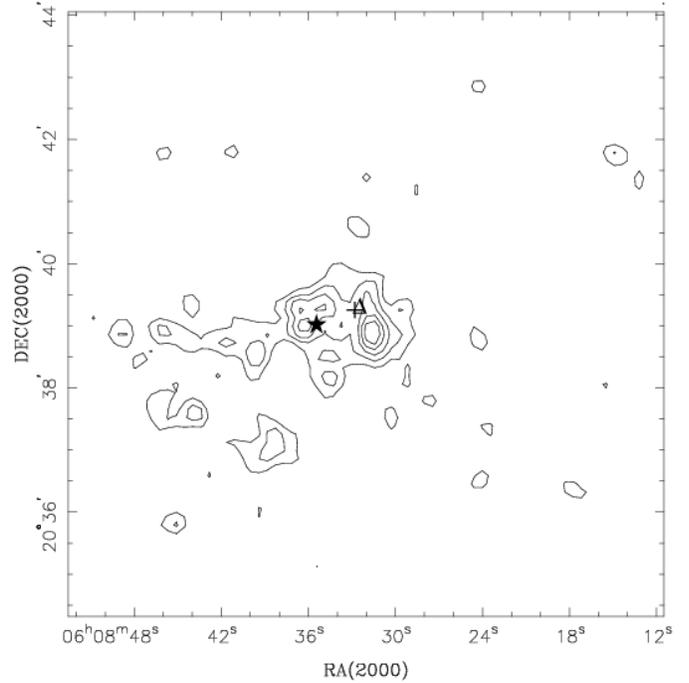}}
\caption{The contour map of the stellar surface number density obtained by
counting stars in a $\rm 10\arcsec \times 10\arcsec$ ( $\sim$ 0.1 $\times$
0.1 pc) grid for the cluster associated with IRAS 06055+2039.
The contours are
from 30 to 140 stars pc$^{-2}$ in steps of 20 stars pc$^{-2}$. The lowest contour is
at the 3$\sigma$ level. The positions of the IRAS point source (plus), the
radio peak (open triangle) and the sub-mm peak (star) are shown in the
figure.}
\label{cl_sdm.fig}
\end{figure}

Figure\,\,\ref{cl_sdm.fig} shows the surface density map of the region around IRAS
06055+2039. We see the presence of two peaks. The prominent peak
which lies to the west coincides spatially with the central bright source. 
The secondary peak, which lies $\rm \sim 50\arcsec$ to the east of the
main peak, coincides well with the
departure seen in the surface density profile from the model
profiles around a radial distance of $\sim$ 0.6 pc.
This secondary peak is situated close to the edge of the
sub-mm peaks presented in Fig.\,\,\ref{jcmt.fig}.
The stellar surface density distribution exhibits a 
centrally condensed-type structure rather than a hierarchical-type structure
(Lada \& Lada 2003). This is consistent with the fact that the derived
radial profile fits reasonably well both to the King's model as well
as the inverse radius model. Hierarchical-type complexes do not follow
any well defined profile. 
Though at a less significant level, centrally condensed clusters have
also been seen to show the presence of structures. For example, Lada \& Lada (1995) show
the presence of satellite subclusters in the outer regions of IC 348.
This is consistent with the double peaked structure that we see for
the cluster associated with IRAS 06055+2039.

\subsubsection{Colour-Colour (CC) and Colour-Magnitude (CM) Diagrams}
\label{cccmd.sect}
\begin{figure}
\resizebox{\hsize}{!}{\includegraphics{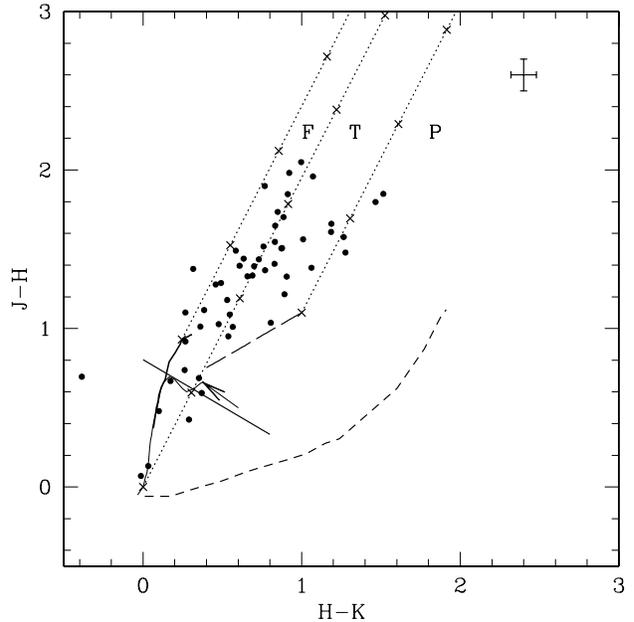}}
\caption{Colour-colour diagram of the infrared cluster
in the IRAS 06055+2039 region.
The two solid curves represent the loci of the main sequence (thin line) and the giant
stars (thicker line) derived from Bessell \& Brett (1988). 
The long-dashed line is the classical T Tauri locus from
Meyer et al. (1997). The parallel dotted lines are reddening 
vectors with the crosses placed along these lines at intervals corresponding to
five magnitudes of visual extinction. We have assumed the interstellar reddening law
of Rieke \& Lebofsky (1985) ($A_{J}/A_{V}$ = 0.282; $A_{H}/A_{V}$ =
0.175 and $A_{K}/A_{V}$ = 0.112). The short-dashed line represents the
locus of the Herbig AeBe stars (Lada \& Adams 1992). The plot 
is classified into three regions namely ``F", ``T" and ``P" (see text
for details).
The colours and the curves shown in the figure are all transformed to the
Bessell \& Brett (1988) system. The solid line shown is drawn
tangential to the turn-off point of the main sequence locus. The arrow points to the position
corresponding to the central
brightest star in the cluster. The mean photometric errors are
shown in the upper right corner.}
\label{cc.fig}
\end{figure}

The ($H-K$) versus ($J-H$) CC diagram for the cluster associated with IRAS 06055+2039
is shown in Fig.\,\,\ref{cc.fig}. In this figure we have plotted 
53 sources which have good quality photometric magnitudes in all the
three $JHK_{s}$ bands (2MASS `read-flag' value of 1 -- 3).
Henceforth, for the analysis of 2MASS data, we will be
using only good quality photometric data which have the above
`read-flag' values.
For clarity we have classified the CC diagram into three regions
(e.g. Sugitani et al. 2002; Ojha et al. 2004a \& b).
The ``F" sources are located within the reddening bands
of the main sequence and the giant stars. These sources are generally
considered to be either field stars, Class III objects or Class II
objects with small NIR excess. ``T" sources populate the region
redward of the ``F" region but blueward of the reddening line 
corresponding to the red end of the T Tauri locus. These sources are classical
T Tauri stars (Class II objects) with large NIR excess or Herbig
AeBe stars with small NIR excess. 
Redward of the ``T" region is the ``P"
region which are mostly protostar-like Class I objects and Herbig AeBe
stars. 
Majority of sources in our sample are almost equally distributed 
in the  ``F" and the ``T" regions, whereas, only four lie in the ``P"
region. A total of 18 out of 80 (22\%) sources show infrared
excess (i.e. sources populating the ``T" and the ``P" regions).
However, it is important to note here that this NIR excess
fraction is just the lower limit as several cluster members detected
in the $K_{s}$ band were not detected in the other two shorter
wavelength bands and hence are not part of this sample. 
The NIR excess in pre-main sequence (PMS) stars is due to the
optically thick circumstellar disks/envelopes. These disks/envelopes
become optically thin with age hence the fraction of NIR excess stars
decreases with age. For very young ($\le$ 1 Myr) embedded clusters the
fraction is $\sim$ 50 \% (Lada et al. 2000; Haisch et al. 2000) and it
decreases to $\sim$ 20 \% for more evolved (2 -- 3 Myr) clusters
(Haisch et al. 2001; Texeira et al. 2004). The fraction of stars with NIR excess 
seen in this cluster suggests an upper limit 2 -- 3 Myr on the age which is
reasonably consistent with that suggested from the core radius
value.  
The age estimates for the cluster NGC 2175 associated
with the extended HII region Sh 252 are 2 Myr (Grasdalen \& Carrasco 1975)
and 1 -- 2 Myr (K\"{o}mpe et al. 1989) which agrees
rather well with our estimates for the cluster associated with IRAS
06055+2039 which also belongs to the Sh 252 complex.

We have calculated the extinction by de-reddening the stars in the CC diagram.
The stars are shifted to a line drawn tangential to the turn-off point
of the main sequence locus (see Fig.\,\ref{cc.fig}). The amount of shift gives an estimate of 
the extinction of individual stars.
The extinction values range from $A_{V}$ $\sim$ 0 to 13 mag with an average
foreground extinction of $A_{V}$ $\sim$ 7 mag. The range of extinction
values obtained shows up as the spread of stars along the reddening
band in the CC diagram. This indicates that the cluster 
is partially embedded (Teixeira et al. 2004). Also according to Lada \&
Lada (2003), low extinction values ($A_{V}$ $\sim$ 1 -- 5 mag) are
typical of partially embedded clusters.

\begin{figure}
\resizebox{\hsize}{!}{\includegraphics{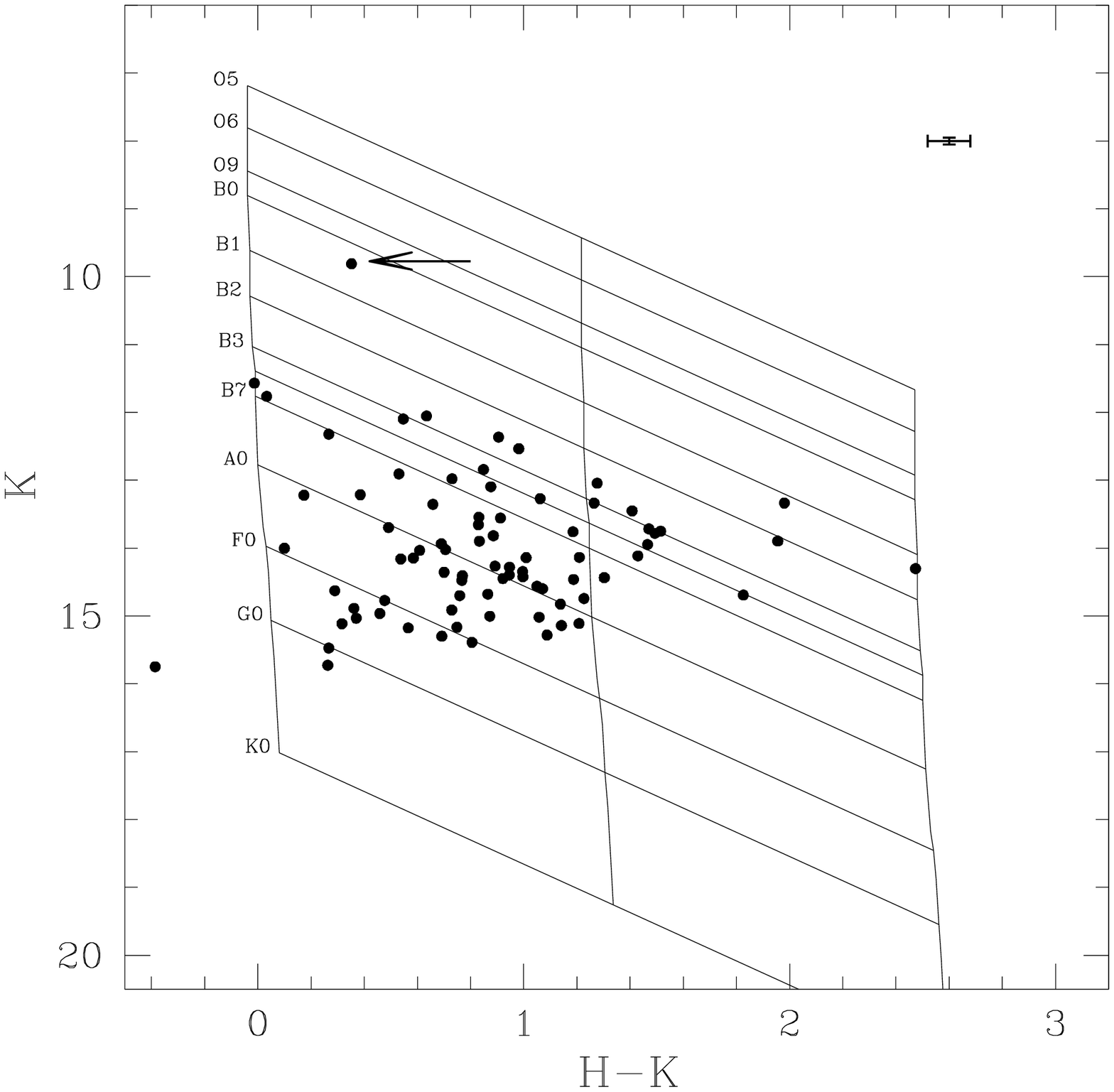}}
\caption{Colour-magnitude diagram of the infrared cluster
in the IRAS 06055+2039 region. The nearly vertical solid lines
represent the zero age main sequence (ZAMS) loci with 0, 20 and 40 magnitudes of visual
extinction corrected for the distance. The slanting lines show the reddening vectors for each spectral
type. The magnitudes and the ZAMS loci are all plotted in the Bessell
\& Brett (1988) system. The arrow points to the position corresponding
to the central brightest star in
the cluster. The mean photometric errors are shown in the upper right
corner.}
\label{cmd.fig}
\end{figure}

Figure\,\,\ref{cmd.fig} shows the ($H-K$) versus $K$ colour-magnitude
(CM) diagram for 79 sources with good quality $HK$ magnitudes. 
Using the zero age main sequence (ZAMS) loci and
the reddening vectors, we estimate the spectral type of the brightest star in the cluster 
to be $\sim$ B0.5. This is the central IRAS point
source and from the NIR estimates seems to be the most massive star in
the cluster. A similar estimate for the spectral type is also obtained
from the analysis of the ($J-H$) versus $J$ CM diagram. 

\subsubsection{$K_{s}$ - Band Luminosity Function}
We use the 2MASS $K_{s}$ - band star counts
to derive the $K_{s}$ - band luminosity function (KLF) for the embedded cluster. 
In order to obtain the KLF of the cluster it is essential to account for
the background and foreground source contamination. For this purpose we use both
the Besan\c{c}on Galactic model of stellar population synthesis (Robin et al. 2003) 
and the observed control field star counts. We use the same control
field as described in Sect\,\,\ref{ssnd.sect}.

\begin{figure*}
\centering
\resizebox{\hsize}{!}{\includegraphics{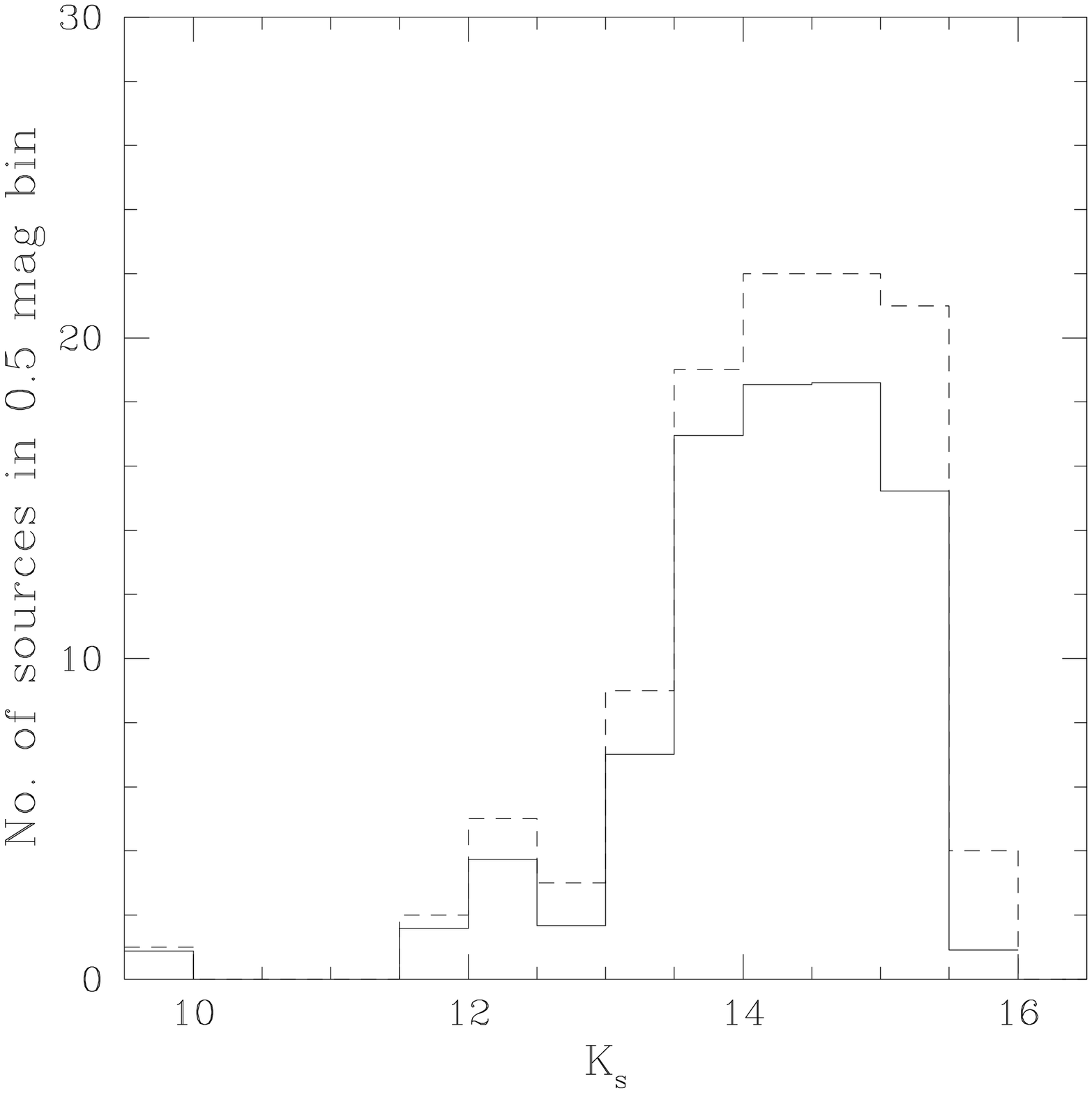} \includegraphics{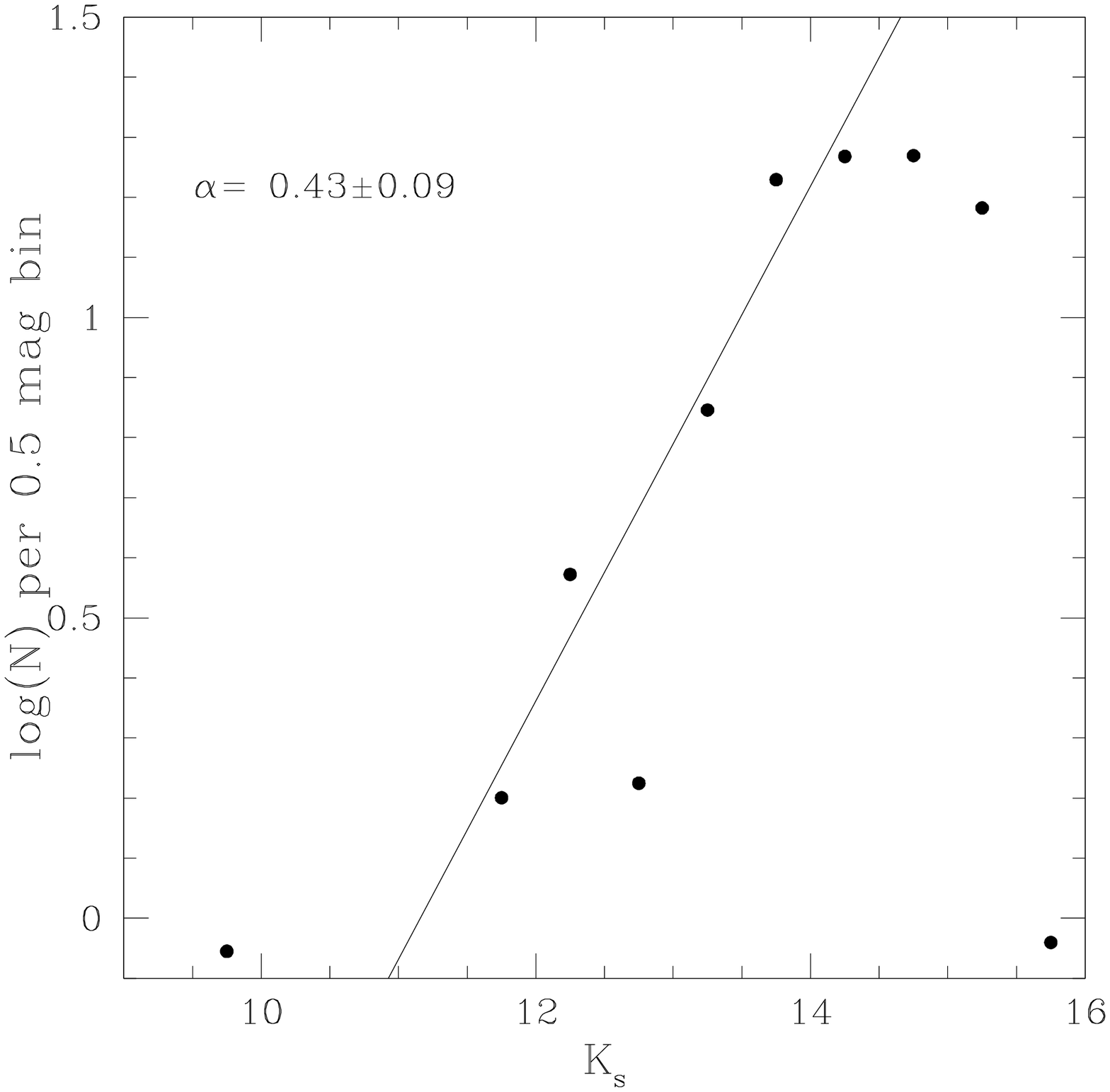}}
\caption{Left: The corrected $K_{s}$ - band luminosity function (KLF)
 for the cluster around IRAS 06055+2039 is shown
(solid line). The dotted line is the luminosity function without
foreground/background correction.
Right: The KLF shown as logN versus the $\rm K_{s}$ magnitude. 
The straight line is the least squares fit to the data points
in the magnitude range 11.5 -- 14.5.}
\label{klf_klf_slope.fig}
\end{figure*}

Star counts were predicted using the Besan\c{c}on model in the
direction of the control field. 
We have checked the validity of the simulated model by comparing the
model KLF with that of the control field and found both the KLFs to
match rather well.
As mentioned in the previous section, the average foreground extinction is determined to be 
A$_{V}$$\sim$ 7 mag. Hence,
assuming spherically symmetric geometry, the background population
is then seen through a cloud with
extinction upto A$_{V}$$\sim$ 14 mag (7 $\times$ 2). Model simulations with A$_{V}$
= 0 mag and d $<$ 2.6 kpc gives the foreground contamination. The background
population is generated with A$_{V}$ = 14 mag and d $>$ 2.6 kpc. We determine the
fraction of the contaminating stars (foreground + background) over the total
model counts. This fraction is used to scale the observed control
field and subsequently the star counts of the modified control field
are subtracted from the KLF of the cluster to obtain the final corrected KLF which
is shown in the left panel of Fig.\,\,\ref{klf_klf_slope.fig}.

The right panel of Fig.\,\,\ref{klf_klf_slope.fig} shows the KLF plotted as $\rm log N$ versus the $K$
magnitude. The KLF can be written as a power-law
\begin{equation} 
\frac{dN(K_{s})}{dK_{s}} \propto 10^{\alpha K_{s}}
\end{equation}
where, the left side of the equation denotes the number of stars per unit
magnitude bin and $\alpha$ is the slope of the power-law. A linear least squares 
fitting algorithm is used to fit the above power-law to
the KLF in the magnitude range $K_{s} = 11.5 \,\rm to \,14.5$. We obtain a value of $\alpha$ =
0.43$\pm$0.09 for the cluster. The least square fitting is done taking
into account the statistical errors on individual data points.

Within the quoted errors the estimate of the power-law slope 
is consistent with the average value of slopes ($\alpha \sim 0.4$) 
obtained for other young clusters (Lada et al. 1991; Lada \& Lada 1995; Lada
\& Lada 2003). 
The power-law slope values obtained for other embedded clusters like
W3 main and NGC 7538 are significantly lower ($\alpha$ $\sim$ 0.17 to
0.33 -- Ojha et al. 2004a \& b; Balog et al. 2004).
However, it should be noted here that these clusters are much younger ($\la$ 1
Myr) and the surveys are deeper ($K_{s}\le 17.5$) which
probe the low mass stellar population down to $\sim \rm
0.1M_{\odot}$.  

\begin{figure}
\resizebox{\hsize}{!}{\includegraphics{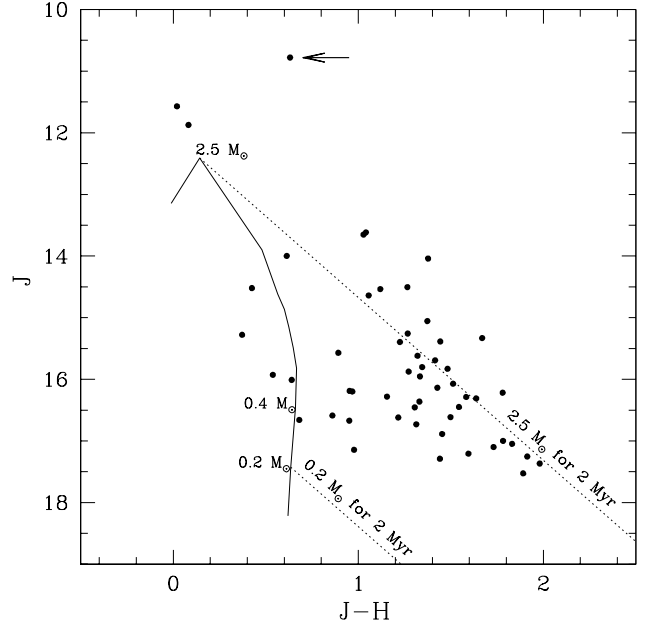}}
\caption{The mass spectrum for the cluster associated with IRAS 06055+2039.
The solid curve represents the model isochrone from Palla \& Stahler (1999) 
for a 2 Myr PMS stellar population for the mass range 0.1 -- 3 $\rm M_{\odot}$.
The slanted dotted lines are the
reddening vectors for 2.5 and 0.2 $\rm M_{\odot}$ PMS stars respectively. 
Sources which lie above the reddening vector for 2.5 $\rm
M_{\odot}$ are luminous massive stars. The
arrow points to the position corresponding to the central massive star
in the cluster.}
\label{mass_spec.fig}
\end{figure}

We estimate the masses of the sources in the cluster by comparing them with the
evolutionary models of Palla \& Stahler (1999). Figure\,\,\ref{mass_spec.fig}
shows the ($J-H$) versus $J$ CM diagram for the cluster field. We use the $J$-band
magnitudes rather than $K_{s}$ because it is less affected by emission
from circumstellar material. The solid curve
represents the ZAMS isochrone for a 2 Myr cluster from Palla \&
Stahler (1999). Majority of sources have a typical mass of 
$\sim \rm 2 M_{\odot}$ which we assume to be representative 
of the stellar population in the cluster. For a log-normal IMF,
the power law slope ($\gamma$), which is 1.35 for a Salpeter IMF,
is variable and is given by $\rm \gamma = 0.94 + 0.94\,log(m_{\star})$,
where $m_{\star}$ is the stellar mass (Miller \& Scalo 1979; Lada et al. 1993). 
Using this relation, we estimate $\gamma \sim 1.2$ for this mean mass.
The power-law slope for the mass to
luminosity relation is $\beta$ $\approx 1$ for clusters of age $\sim 10^{6}$
Myr (Simon et al. 1992; Lada et al. 1993). The corresponding slope of
the KLF,
$\alpha$ ($=\gamma/2.5\beta$), is 0.48 which is consistent with the
value obtained from the least squares fit to the KLF.
As is seen from Fig.\,\ref{mass_spec.fig}, the mass of the majority of
sources in the cluster
are below 2.5 $\rm M_{\odot}$ and the lowest mass limit is $\sim$ 0.4 $\rm
M_{\odot}$ from our sample.

%--------------------
\subsection{Spatial Distribution of UIBs from MSX Data}
\label{uib.sect}
\begin{figure}
\resizebox{\hsize}{!}{\includegraphics{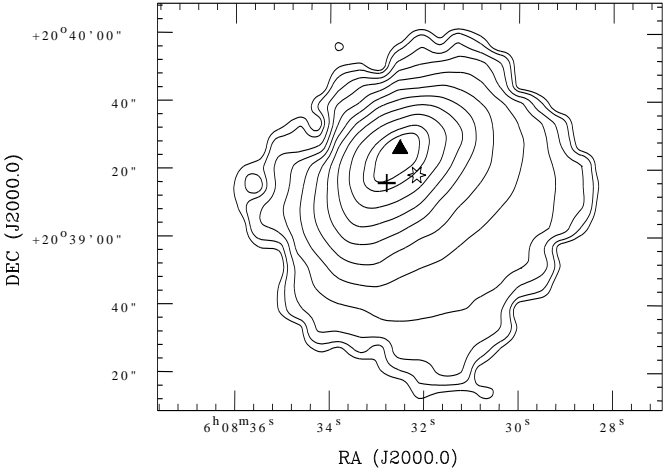}}
\caption{Spatial distribution of the total radiance in the UIBs for the region around IRAS
06055+2039 as extracted from the MSX four band images. The contour
levels are at 5, 10, 20, 30, 40, 50, 60, 70, 80, 90 and 95 \% of 
the peak emission of $\rm 4.1\times10^{-5}$ W\,m$^{-2}$\,Sr$^{-1}$. The
peak position of the UIB distribution (filled triangle), the 1280
MHz radio emission (open star) and the IRAS point source position
(plus) are also indicated.}
\label{pah.fig}
\end{figure}

We have used the scheme developed by Ghosh \& Ojha (2002) to extract the contribution of
UIBs (due to the Polycyclic Aromatic Hydrocarbons (PAHs)) from the mid-infrared images in
the four MSX bands. 
The emission from each pixel is assumed to be a combination of two
components. The first is the thermal continuum from dust grains (gray
body) and the second is the emission from the UIB features falling
within the MSX bands. The scheme
assumes the dust emissivity to follow the power law of the form 
$\epsilon_{\lambda} \propto \lambda^{-1}$ and the
total radiance due to the UIBs in band C to be proportional to that in
band A. A self consistent non-linear chi-square minimization technique
is used to estimate the total emission from the UIBs, the temperature
and the optical depth.
The spatial distribution of emission in the UIBs
with an angular resolution of $\rm \sim 18\arcsec$ (for the MSX survey) extracted
for the region around IRAS 06055+2039 is shown in Fig.\,\,\ref{pah.fig}. 

Comparing the morphology with the radio continuum maps, we see that
the emission from the UIBs is much more extended though the gross
morphologies are similar with a relatively steep intensity gradient towards the N-E and a
smoothly decreasing intensity distribution to the S-W. The peak
position of the UIB distribution matches rather well with the radio
peak.

The integrated emission from the region around IRAS 06055+2039
in the UIB features within the band A of MSX (viz., 6.2, 7.7
\& 8.6\,$\mu$m) is found to be 2.85 $\times 10^{-12}$ Wm$^{-2}$
(see Fig.\,\,\ref{pah.fig}). For comparison, we have estimated the emission in individual
UIB features from the IRAS LRS spectrum (covering 8 -- 22\,$\mu$m),
for this source. The total emission in the 7.7 and 8.6 $\mu$m features is $\sim$
7.33 $\times 10^{-12}$ Wm$^{-2}$, in the 11.3\,$\mu$m feature is $\sim$
1.32 $\times 10^{-12}$ Wm$^{-2}$ and an upper limit for the 12.7\,$\mu$m
feature is 5.9 $\times 10^{-13}$ Wm$^{-2}$. The last value is an
upper limit due to possible contamination from the Ne [II] line
at 12.8\,$\mu$m. Hence, the UIB emission extracted from the MSX band A
is $\la$ 39\% of the estimate from LRS. This is reasonable considering
the larger effective field of view for the latter.

%------------------------------------------------------------------------------------------
\subsection{Emission from Ionized Gas}
The radio continuum emission from the ionized gas associated with the region
around IRAS 06055+2039 at 1280 and 610 MHz is shown in
Fig.\,\,\ref{1280_610.fig}. The integrated flux densities from these maps are
183 and 282 mJy at 1280 and 610 MHz
respectively. It should be noted here that the flux densities are
obtained by integrating upto 3\,$\sigma$ level, where $\sigma$ is the
rms noise of the maps. The integrated flux densities obtained from the
GMRT maps are consistent with the results of Felli et al. (1977) and White \& Gee
(1986). Felli et al. (1977) get values of 230 and 205 mJy at 1415 and
4995 MHz respectively. White \& Gee (1986) estimate the total
integrated flux density at 5 GHz to be 198 mJy. 
The contour maps display a cometary morphology with a bright arc-shaped edge on the N-E side and a smoothly
decreasing intensity distribution on the opposite side. The cometary
morphology is more clearly seen in the 610 MHz map. 
Such a morphology implies that the HII region
is ionization bounded towards the N-E and density bounded towards the
S-W. The position of the brightest radio peak matches
with the central bright star seen in the infrared images.
Similar morphology is seen from the 4995 MHz map of Felli et al. (1977). They
report the presence of several dense clumps. 
According to them the radio continuum emission from this HII
region (S252 A) is spatially separate from the larger extended HII region associated with Sh
252. CO line observations of Lada \& Wooden (1979) show that this
compact HII region (S252 A), the associated H$_{2}$O maser and CO bright spot are
nearly coincident and located near the interface of the molecular cloud with the extended S252 HII
region. The detection of the H$_{2}$O maser also implies the relative youth of the region. 
Lada \& Wooden (1979) also suggest S252 A to be in the early
stages of stellar evolution in which a massive star (or stars) has
reached the main sequence and created a compact HII region within its parental molecular cloud.
VLA maps at 15 and 5 GHz of White \& Gee (1986) also trace similar cometary morphologies.

Using the low frequency flux densities at 1280 and 610 MHz from our GMRT
observations and 5 GHz data from White \& Gee (1986), we derive
the physical properties of the compact core of the HII 
region associated with IRAS 06055+2039. Mezger \& Henderson (1967)
have shown that for a homogeneous and spherically
symmetric core, the flux density can be written as
\begin{equation}
S = 3.07 \times 10^{-2} T_{e} \nu^{2} \Omega (1 - e^{-\tau(\nu)})
\end{equation}
where,
\begin{equation}
\tau(\nu) = 1.643a \times 10^{5} \nu^{-2.1} (EM) T_{e}^{-1.35}
\end{equation}
where, $S$ is the integrated flux density
in Jy, $T_{e}$ is the electron temperature in K, $\nu$ is the
frequency of observation in MHz, $\tau$ is the optical depth, $\Omega$
is the solid angle subtended by the source in steradians and $EM$ is
the emission measure in $\rm cm^{-6} pc$. $a$ is a correction factor
and we use a value of 0.99 (using Table 6 of Mezger \& Henderson 1967)
for the frequency range 0.6 -- 5 GHz and $T_{e} = \rm 8000 K$. 
The two GMRT maps are convolved to a common angular resolution of $\rm
12\arcsec \times 12\arcsec$ which is the resolution
of the 5 GHz map of White \& Gee (1986).
In our case since the core is unresolved, $\Omega$ is taken as 
this synthesized beam size (i.e $\Omega = 1.133 \times \theta_{a} \times
\theta_{b}$, where $\theta_{a}$ and $\theta_{b}$ are the half power
beam sizes). The peak flux densities of the core in
the 0.6 -- 5 GHz frequency range appear to lie in the optically
thin region. Using these peak flux densities we derive the emission measure for an estimated
electron temperature.

The electron temperature ($ T_{e}$) of HII regions is known to
increase linearly with Galacto-centric distance ($ D_{G}$) (Deharveng
et al 2000 and references therein). This is due to the decrease in heavy element
abundance with $ D_{G}$ which results in higher $ T_{e}$.
The values of $ T_{e}$ derived by Omar et al. (2002) for a
sample of three Galactic HII regions are also consistent with the
relationship given in Deharveng et al. (2000). Assuming $
D_{G}$ as 10 kpc (Shirley et al. 2003) for IRAS 06055+2039 (S252A), we
obtain a value of $\sim$ 8000 K for the electron temperature.
Using this value of the electron temperature, the peak flux densities
were used to fit the above equations (Eqns. 3 \& 4). The best fit value for
the emission measure is $\rm 8.8 \pm 0.4 \times 10^{4} cm^{-6} pc$.
We get an estimate of $\rm 1.05 \times 10^{3} cm^{-3}$ for the electron density ($n_{e}$ $=
(EM/r)^{0.5}$, $r$ being the core size which in this case corresponds
to the synthesized beam size). 
These values agree reasonably well with the estimates of Felli et al.
(1977) for the brightest peak (A3) of the component S252 A (see Fig. 5
of Felli et al. 1977). They
derive a value of $\rm 9.9 \times 10^{4} cm^{-6} pc$ and $\rm 5.95 \times 10^{2}
cm^{-3}$ for $EM$ and $n_{e}$ respectively. They assumed an electron
temperature of $\rm 10^{4} K$ and a distance of 2 kpc. It should also be
noted here that the 5 GHz map of Felli et al. (1977) has a larger beam
size ($\rm \sim 8\arcsec\times21\arcsec$).

Taking the total integrated flux density of 183 mJy at 1280 MHz and using
the formulation of Schraml \& Mezger (1969) and the table from 
Panagia (1973; Table II), we estimate the exciting star of this HII
region to be of the
spectral type B0 -- B0.5. This is consistent with the spectral class
obtained by Felli et al. (1977) and White \& Gee (1986).
The FIR flux densities from the IRAS PSC yield a luminosity of
$\sim$ 10$^{4}\rm L_{\odot}$ which implies an exciting star
of spectral type B0.5, in good agreement with radio measurements.

In addition to the diffuse emission seen in our
1280 MHz map, we also detect a few discrete sources probably
representing high density clumps which are listed in 
Table\,\,\ref{radio_sources.tab}. We designate them
as R1, R2,....R7 and their positions are marked in
Fig.\,\ref{1280_610.fig}. Three such dense clumps were also detected
in the 5 GHz map of Felli et al. (1977). The position of R1 is spatially coincident
with the position of the central bright and
massive ($\sim$ B0.5) star seen in the infrared cluster and is
possibly the exciting source of the HII region. The other dense clumps
could also be possible discrete radio sources but the resolution of
our map makes it difficult to comment on their nature.
\begin{table}
\caption{Discrete sources extracted from the 1280 MHz map of the
region associated with IRAS 06055+2039.}
\label{radio_sources.tab}
\begin{tabular}{c|c|c|c}
\hline\hline
Source     & RA (2000.0)   & DEC (2000.0)   & Peak Flux density \\ 
              & (h m s)       & (d m s)        & (mJy/beam)  \\
\hline
R1             & 06 08 32.16    & +20 39 18.7    & 5.7       \\
R2             & 06 08 31.95    & +20 39 23.1    & 5.4 	     \\
R3             & 06 08 32.79    & +20 39 16.1    & 4.3       \\
R4             & 06 08 31.04    & +20 39 24.9    & 3.3       \\
R5             & 06 08 30.92    & +20 39 30.3    & 2.8       \\
R6             & 06 08 31.39    & +20 39 02.1    & 3.2       \\
R7             & 06 08 32.20    & +20 39 01.7    & 2.8       \\
\hline
\end{tabular}
\end{table}
These seven clumps contribute $\sim$ 5\% of the total integrated
emission from the ionized region around IRAS 06055+2039, the remaining
being of diffuse nature. 

\subsection{Emission from Shocked Neutral Gas}
\begin{figure*}
\resizebox{\hsize}{!}{\includegraphics{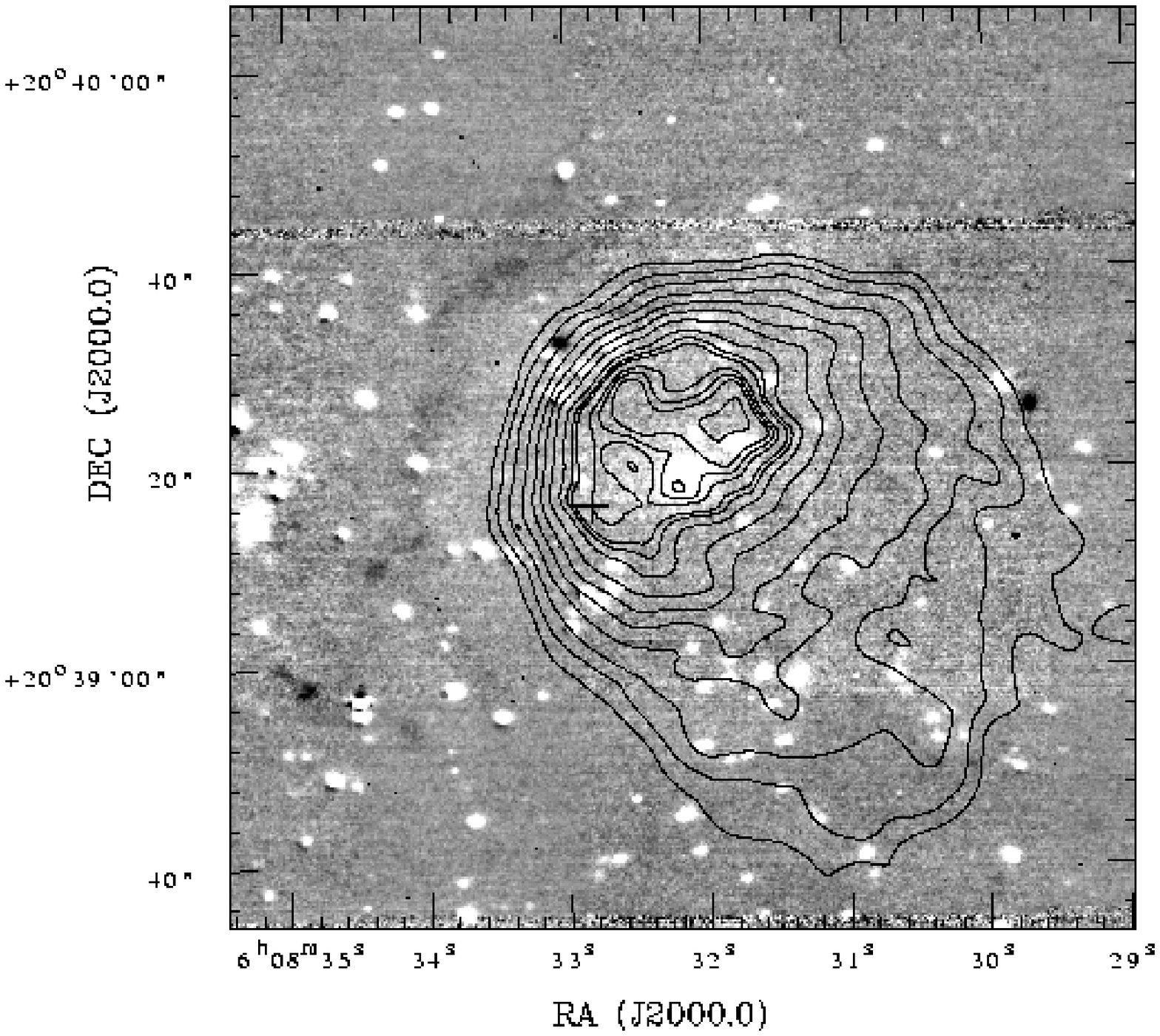} \includegraphics{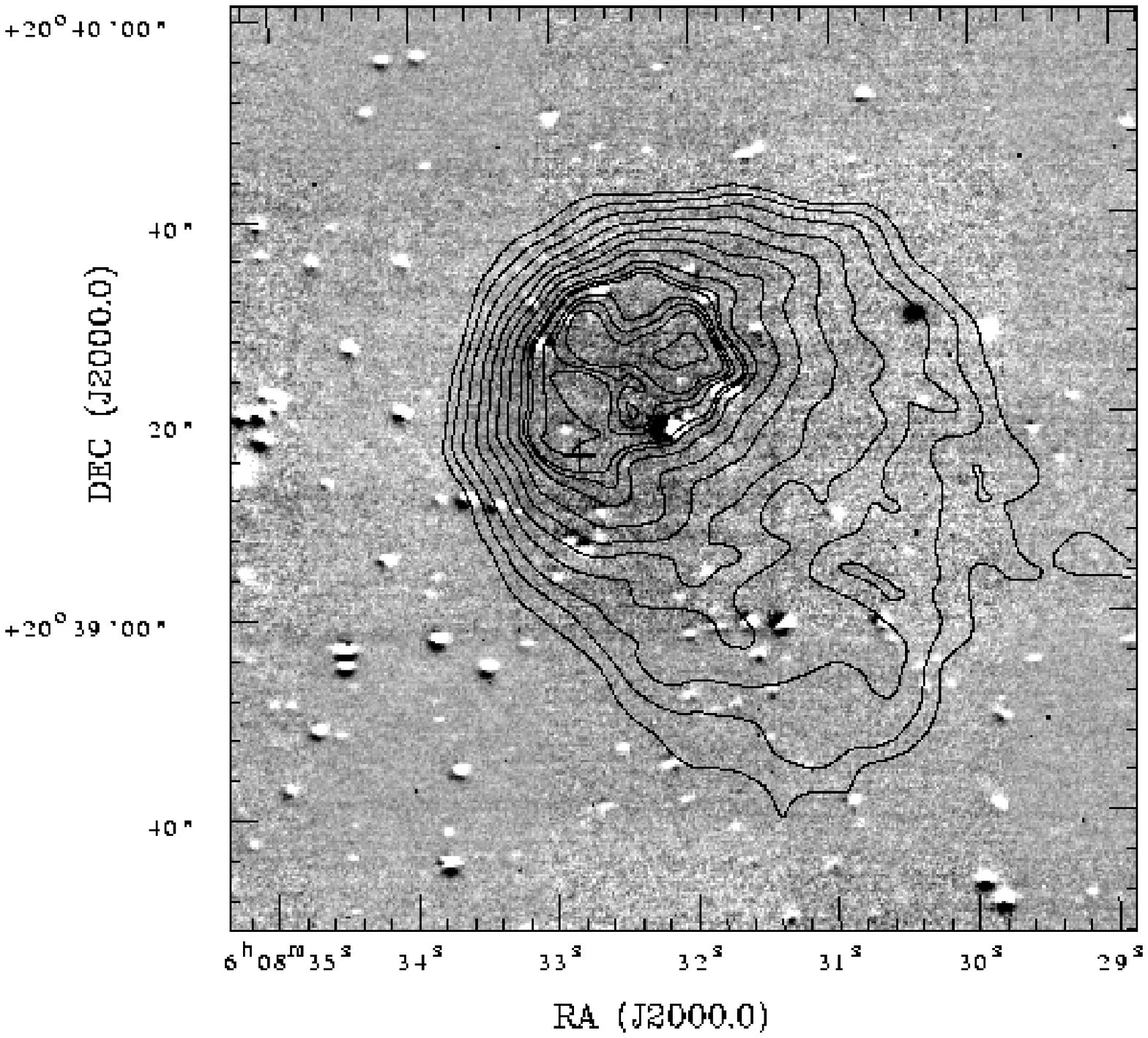}}
\caption{610 MHz radio contours overlaid over the continuum subtracted
H$_{2}$ (left) and Br$\gamma$ (right) images for the region associated
with IRAS 06055+2039. The contour levels are same as in
Fig.\,\,\ref{1280_610.fig}. The plus sign in both the images
marks the position of the IRAS point source.}
\label{h2_brg_rad.fig}
\end{figure*}

The rotational-vibrational line of molecular hydrogen (H$_{2}$ (1-0)S1,
2.12\,$\mu$m) traces the shocked neutral gas at the interface between 
the ionized and the molecular gas. In the photo-dissociation regions (PDRs), the
molecular emission traces the first neutral layer beyond the ionization front.
In Fig.\,\,\ref{h2_brg_rad.fig}, we compare the morphologies of the 610 MHz continuum map with the
continuum subtracted narrow-band H$_{2}$ (left) and Br$\gamma$ (right) images. 
It is interesting to note that the shocked molecular
hydrogen envelopes the radio emitting region. The
Br$\gamma$ image shows the presence of a faint diffuse emission
which correlates well with the
cometary morphology of the radio continuum emission. The H$_{2}$ arc which traces the
ionization front lies beyond the Br$\gamma$ emission. Comparison with
the 1280 MHz continuum map also shows similar morphology. From the position of
the H$_{2}$ arc, we estimate the radius of the HII region to be $\sim$ 0.4 pc.

%---------------------------------------------------------------------------
\subsection{Emission from Dust: Temperature, Optical Depth and Dust Mass}
\begin{figure*}
\centering
\resizebox{\hsize}{!}{\includegraphics{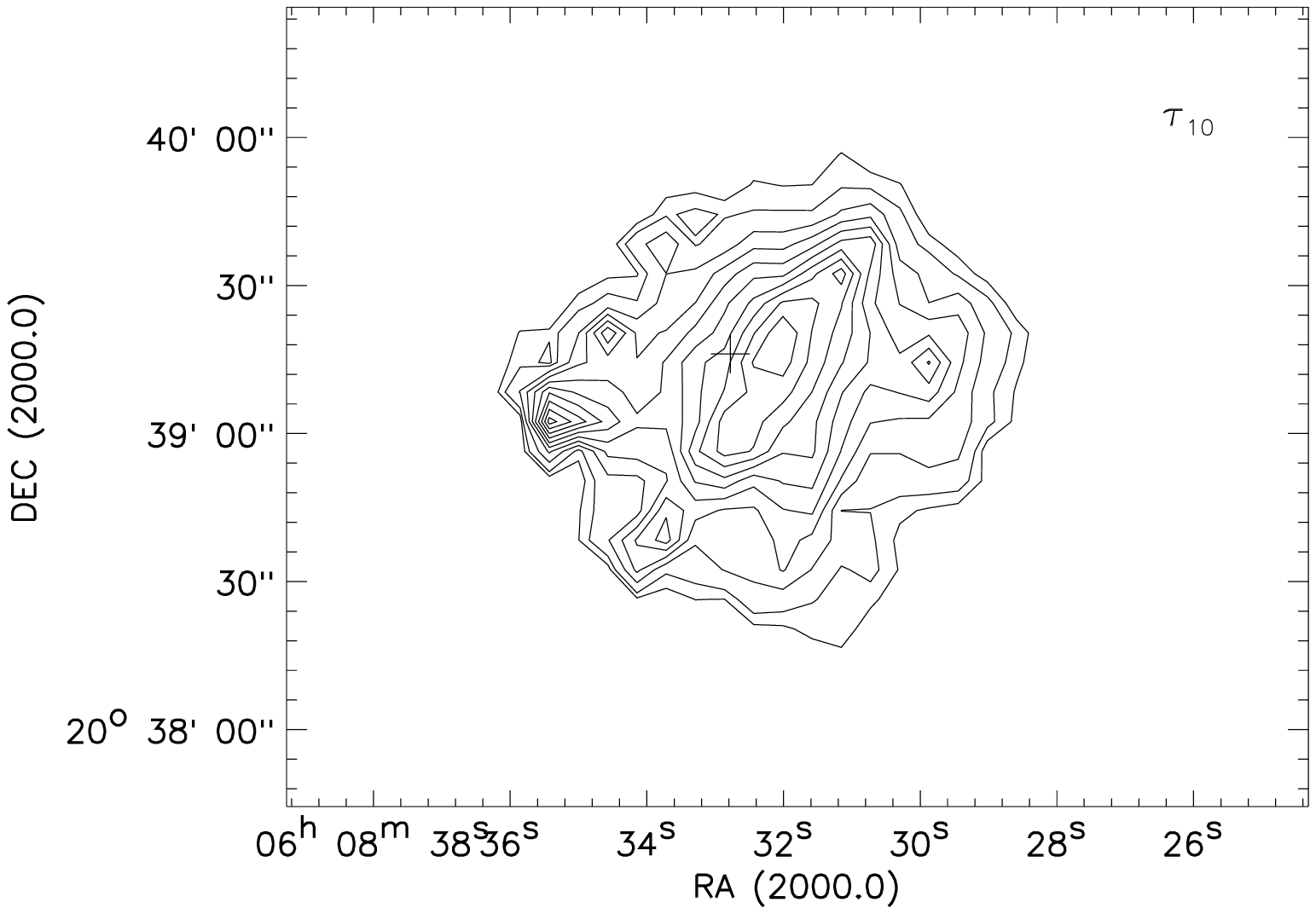}\includegraphics{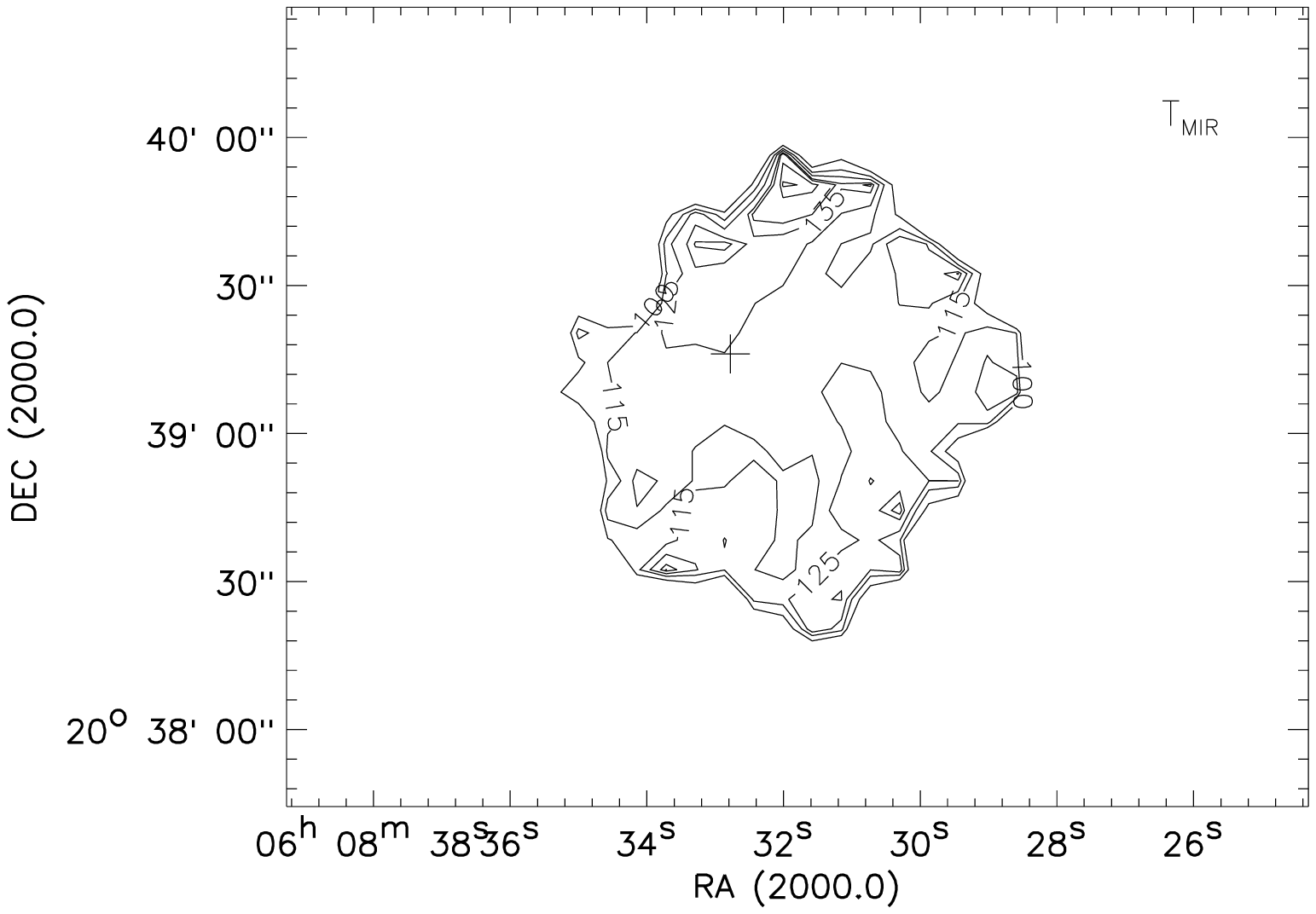}}
\caption{Left: The spatial distribution of dust optical depth $\tau_{10}$
in the region around IRAS 06055+2029. The contour levels are at
5, 10, 20, 30, 40, 50, 60, 70, 80 and 90 \% of the peak value of
$\rm 1.4 \times 10^{-4}$.
Right: The spatial distribution of the mid-infrared dust temperature
in this region. 
The contour levels are for the 100,115,125,135,140,150 and 154
K. 
The peak temperature is 155 K. The plus sign in both the images
marks the position of the IRAS point source.}
\label{msx_tautemp.fig}
\end{figure*}

As discussed in Sect.\,\,\ref{uib.sect}, the MSX images were 
used to obtain the spatial distribution of temperature and optical
depth ($\tau_{10}$) of warm dust with the assumption that the dust is optically
thin and the dust emissivity follows the power law of the form
$\epsilon_{\lambda} \propto \lambda^{-1}$ (Mathis et al. 1983;
Scoville \& Kwan 1976).
We obtain peak values of $1.4\times10^{-4}$ and 155 K for $\tau_{10}$
and the dust temperature respectively.
Figure\,\,\ref{msx_tautemp.fig} shows the optical depth and the
mid-infrared dust temperature maps.
Morphologically, the spatial distribution of the UIBs and the optical depth contours are similar
with the intensity peaks matching rather well. This indicates the
presence of higher dust densities near the
embedded cluster. 
The temperature distribution shows lower values near the centre
with a plateau like feature running diagonally along the S-E and N-W
direction. Higher temperatures are towards the periphery with the peak
seen towards the north. The optical depth and temperature are inherently
anti-correlated. Hence, we see that the region has higher optical depth at
the centre with the values decreasing outwards. 
Several peaks seen in the $\tau_{10}$ map could
possibly indicate the clumpy nature of the region. As will be 
evident later, we see similar trends for the
$T\rm(12/25)$ and $\tau_{25}$ maps derived from the IRAS-HIRES
images.

The MSX Point Source Catalog (MSX PSC) lists two mid-infrared sources which
fall within the radio nebulosity of IRAS 06055+2039. We designate them
as M1 and M2.
The MSX PSC flux densities for these two sources are listed in Table\,\,\ref{msx_psc.tab}.
\begin{table}
\caption{Flux densities for the MSX point sources possibly associated with IRAS
06055+2039}
\label{msx_psc.tab}
\begin{tabular}{c|c c}
\hline\hline
                    & \multicolumn{2}{|c}{Flux density (Jy)}\\
\hline
Wavelength ($\mu$m) & M1$^{1}$        & M2$^{2}$	\\
\hline
8.3                 & 3.52            & 0.97               \\
12.1                & 6.24            & 1.40                \\
14.7                & 2.42            & 4.19                \\
21.3                & 6.66            & 13.98              \\
\hline
\end{tabular}

{\tiny $^{1}$G189.7672+00.3407 ($\alpha_{2000}=
06^{h}08^{m}33.02^{s}$; $\delta_{2000} =
20^{\circ}39\arcmin32.04\arcsec$)\\
$^{2}$G189.7677+00.3376 ($\alpha_{2000}= 06^{h}08^{m}32.40^{s}$;
$\delta_{2000} = 20^{\circ}39\arcmin25.20\arcsec$)\\}
\end{table}
Comparing the MSX mid-infrared colours $\rm F_{21/8}$, $\rm F_{14/12}$, $\rm F_{14/8}$
and $\rm F_{21/14}$ of these two sources with study of the Galactic
plane population by Lumsden et al.
(2002), we infer that M1 is possibly a
Herbig AeBe or a foreground star, whereas, M2 falls in the zone
occupied mostly by compact HII regions. The NIR counterpart from 2MASS
catalog for M2 is the central bright IRAS point source. The mid-
and near-infrared colours $\rm F_{21/8}$, $\rm F_{8/K}$, $\rm
F_{21/12}$ and $\rm F_{K/J}$ of this source are also consistent with
compact HII regions. 

\begin{figure*}
\centering
\resizebox{\hsize}{!}{\includegraphics{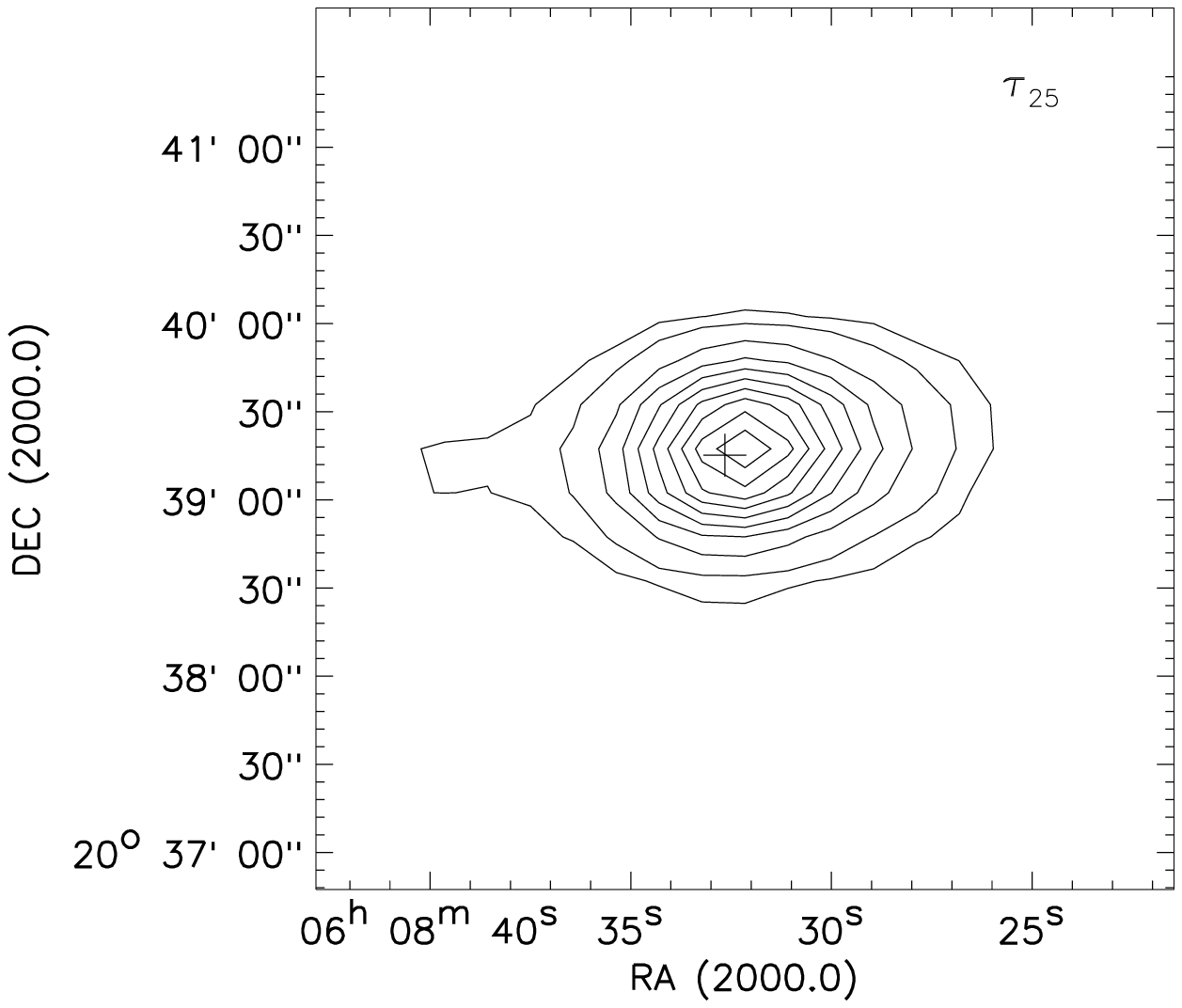} \hskip -2 cm\includegraphics{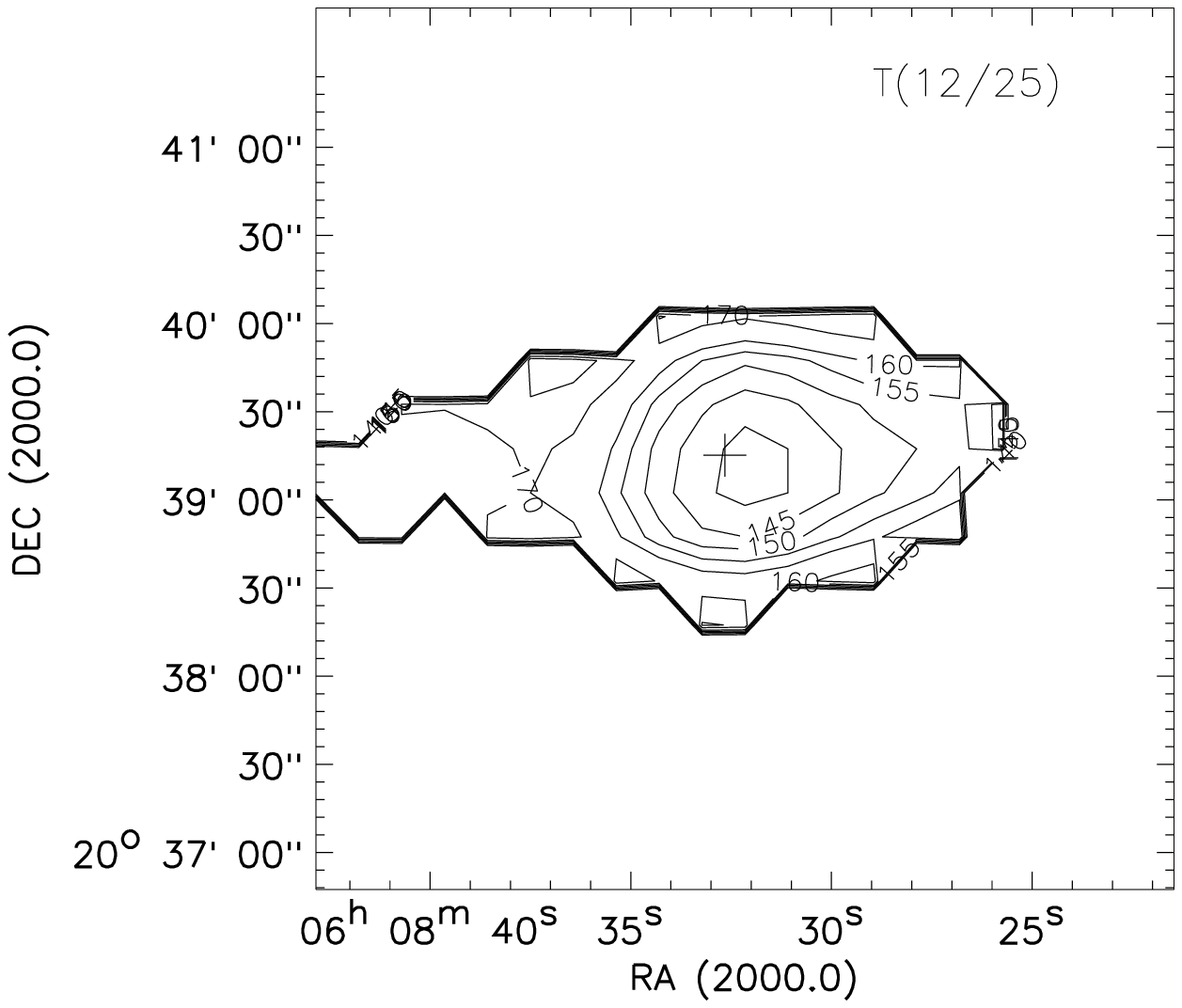}}
\resizebox{\hsize}{!}{\includegraphics{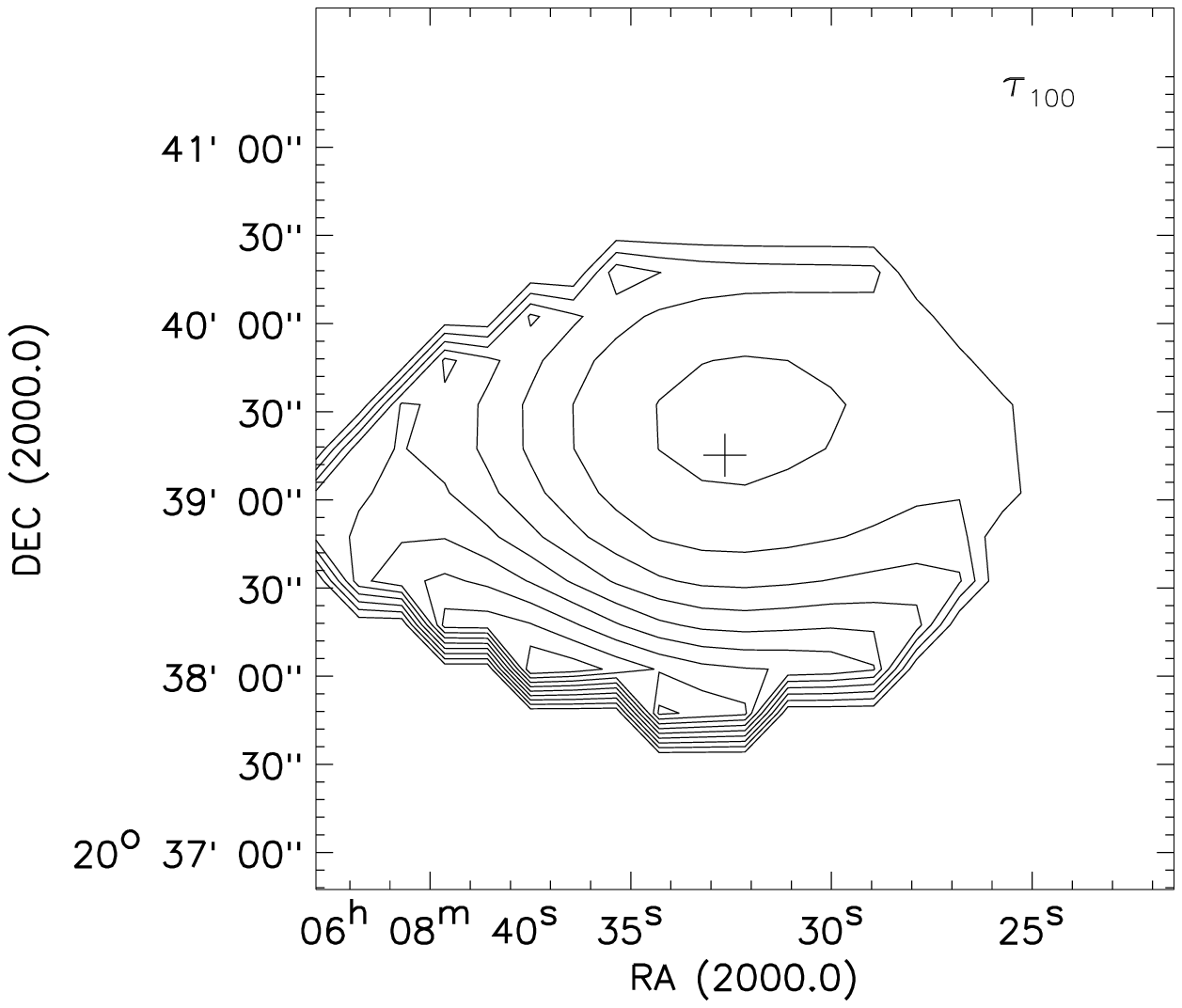} \hskip -2 cm\includegraphics{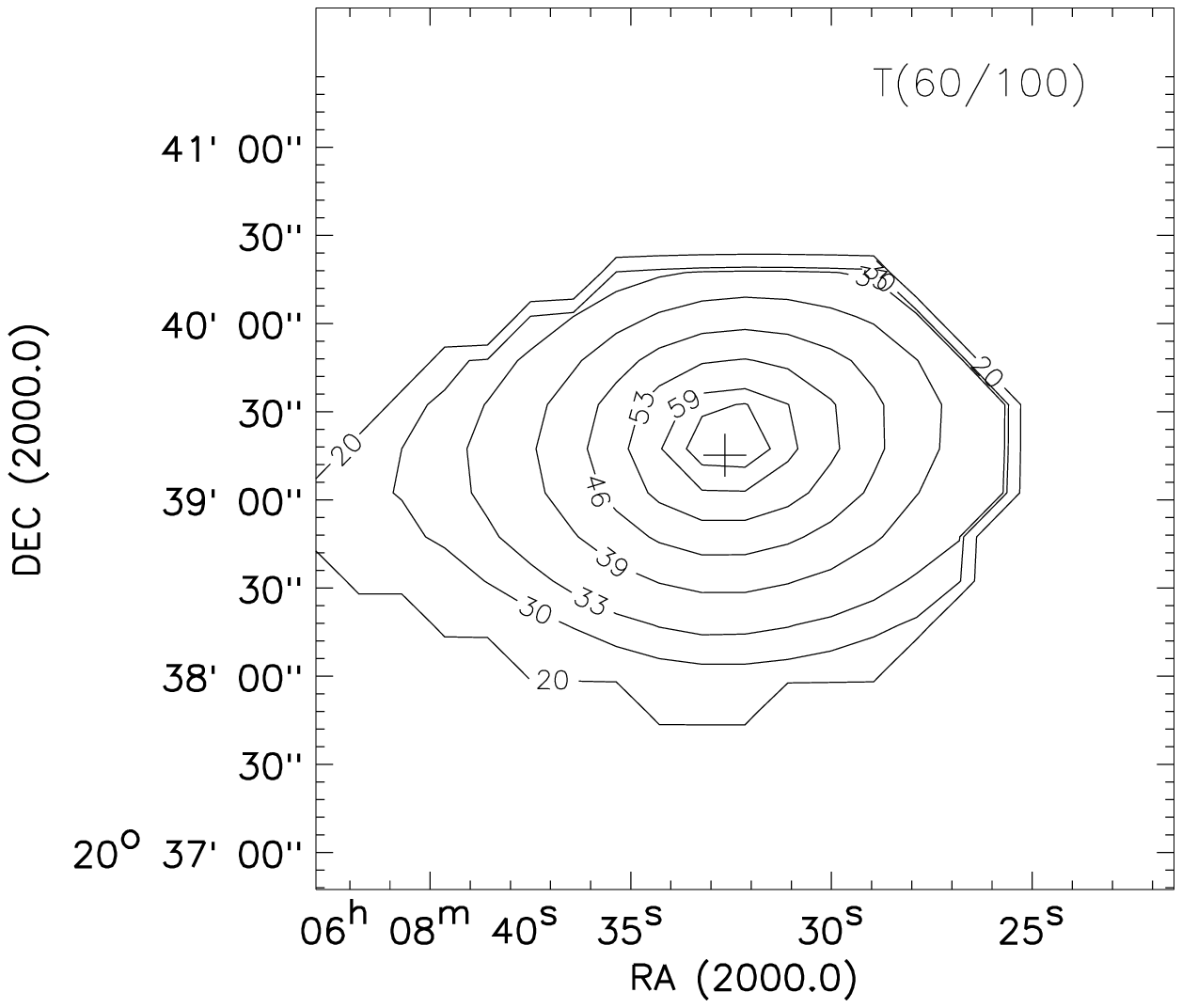}}
\caption{Upper panel -- The dust optical depth ($\tau_{25}$) (left) and colour temperature
($T\rm(12/25)$) (right) maps of the region around IRAS 06055+2039. The optical
depth contours are 5, 10, 20, 30, 40, 50, 60, 70, 80, 90 \% of the peak value of
$2.2\times10^{-5}$. The dust temperature contour levels are at 140,
145, 150, 155, 160, 170 and 180 K from the centre to the periphery.
The peak value is 189 K.
Lower panel -- The dust optical depth ($\tau_{100}$) (left) and colour temperature
($T\rm(60/100)$) (right) maps of the region around IRAS 06055+2039. The optical
depth contours are 10, 20, 30, 40, 50, 60, 70, 80, 90 \% of the peak value of
$1.2\times10^{-2}$ from the centre to the periphery. The dust temperature contour levels 
are at 20, 30, 33, 39, 46, 53, 59 and 62 K. The peak value is 66 K.
The plus sign in all the images marks the position of the IRAS
point source.}
\label{hires_1234.fig}
\end{figure*}

The IRAS-HIRES maps (at 12, 25, 60 and 100\,$\mu$m) were also used to obtain the spatial
distribution of warm and cold dust colour temperatures ($T\rm
(12/25)$, $T\rm (60/100)$) and optical depths ($\tau_{25}$, $\tau_{100}$). 
We assume the dust emissivity to follow the power law of the form
$\epsilon_{\lambda} \propto \lambda^{-1}$.
Figure\,\,\ref{hires_1234.fig} shows the dust temperature and
optical depth maps.
The maps for the optical depth $\tau_{25}$ and colour temperature
$T\rm (12/25)$ represent the warmer dust component. The distribution
is centrally dense with the optical depth peak and hence
lower derived temperature at the centre. This is similar to the distribution
seen in Fig.\,\,\ref{msx_tautemp.fig}. On the other hand $\tau_{100}$
and $T\rm (60/100)$ distributions are from a relatively colder component
which probably forms an envelope around the warmer dust.
Unlike the warmer dust temperature distributions
(obtained from mid-infrared emission), the $T\rm (60/100)$
distribution has its peak at the centre with the temperature
decreasing towards the periphery.
For $\tau_{25}$ and $\tau_{100}$, we obtain peak values of $2.2\times10^{-5}$ and
$1.2\times10^{-2}$ respectively. The dust temperature distributions for 
$T\rm(12/25)$ and $T\rm(60/100)$ peak at 189 and 66 K respectively.
Schreyer et al. (1996) derive a value of 31.2 K for $T\rm(60/100)$ and
$1.21\times10^{-3}$ for $\tau_{100}$ from the IRAS PSC flux densities. This
difference is due to the different resolution of the raw IRAS and the
HIRES processed maps. Scaling the
$\tau_{25}$ peak value, we obtain a value of
$\tau_{10}$ $\sim$ $5.6\times10^{-5}$. 
The difference between this value and that obtained from the MSX data 
could be a result of different beam sizes and/or an inhomogeneous medium. 
Using the peak value of $\tau_{100}$ distribution,
derived from the IRAS-HIRES maps, we
estimate the warm dust mass to be $\sim$ 6 $\rm M_{\odot}$.

We use the emission at submillimetre wavebands 
to study the cold dust environment in the region around IRAS
06055+2039. 
The spatial distribution of the sub-mm emission is shown
in Fig.\,\,\ref{jcmt.fig}. The angular resolutions are $\rm
7\arcsec.8$ and $\rm 15\arcsec.2$ for the 450 and 850\,$\mu$m 
wave bands respectively. The main source (central dense
core which covers the region upto $\sim$ 25\% of the peak intensity)
seems elongated in both the maps. Apart from this dense core, 
several dust clumps are seen which
could have probably formed due to the fragmentation of the original
cloud.

The dust mass can be estimated from the following relation:
\begin{equation}
M_{dust} = 1.88 \times 10^{-4} \left(\frac{1200}{\nu}\right)^{3+\beta} S_{\nu}(e^{0.048\nu/T_{d}} - 1) d^{2}
\end{equation}
This is taken from Sandell (2000) and is a simplified version of Eqn. 6 of
Hildebrand (1983). The above equation assumes the standard Hildebrand
opacities (i.e. $\rm \kappa_{1200GHz} = 0.1 cm^{2}g^{-1}$). 
Here, $S_{\nu}$ is the flux density at frequency $\nu$,
$T_{d}$ is the dust temperature which we assume to be 20 K (Klein et
al. 2005; Mueller et al. 2002), $\beta$ is
the dust emissivity index and is taken to be
2 (Hildebrand 1983) and $d$ is the distance to the source in
kpc. The flux densities are obtained from the JCMT-SCUBA maps shown in
Fig.\,\,\ref{jcmt.fig}. To obtain the flux density of the entire cloud, we have integrated upto the
last contour (which is at 5\% of the peak value).
Using the above relation, we estimate dust masses 
of $\sim$ 70 and 90 $\rm M_{\odot}$
from the 450 and 850\,$\mu$m maps respectively. 
Assuming a gas-to-dust ratio of 100, the above values
translate to total masses of 7000 and 9000 $\rm M_{\odot}$
for the cloud from the 450 and 850\,$\mu$m maps respectively.
We also estimate the total
mass of only the central dense core to be $\sim$ 875 and 1250
$\rm M_{\odot}$ from 450 and 850\,$\mu$m maps respectively. 
This source (S252A) has also been studied by Mueller et al. (2002) at
350\,$\mu$m and more recently by Klein et al. (2005) at 850 and
1300\,$\mu$m. 
Scaling to the distance and gas-to-dust ratio assumed by us, the
corresponding mass from Mueller et al. (2002) is $\sim$
600 $\rm M_{\odot}$ and from Klein et al. (2005) is $\sim$ 100 $\rm
M_{\odot}$. Comparing the mass derived from the 850\,$\mu$m map,
our estimate is very close to
the mass obtained by Mueller et al. (2002) considering the fact that they have
assumed a higher dust temperature (29 K). The mass estimate from the
450\,$\mu$m maps is $\sim$ 35 \% lower which could have been affected
by the large atmospheric extinction correction applied to the data. The mass derived by
Klein et al. (2005) are relatively lower. This could be possibly
due to the fact that the flux density values presented by Klein et
al. (2005) are lower compared to our values.
From the CS line maps, Zinchenko et al. (1998) derive 
a cloud mass of 3132 $\rm M_{\odot}$. It should be noted here that
the CS maps are from a larger region.
Furthermore, a point worth discussing here is the effect of
varying $T_{d}$ and $\beta$. Exploring the range of $T_{d}$ (20 -- 40
K) and $\beta$ (1 -- 2), we infer that the mass estimates can 
vary by upto a factor of $\sim$ 8.

\subsection{Comparison of the Different Components
Associated with IRAS 06055+2039}
\begin{figure}
\resizebox{\hsize}{!}{\includegraphics{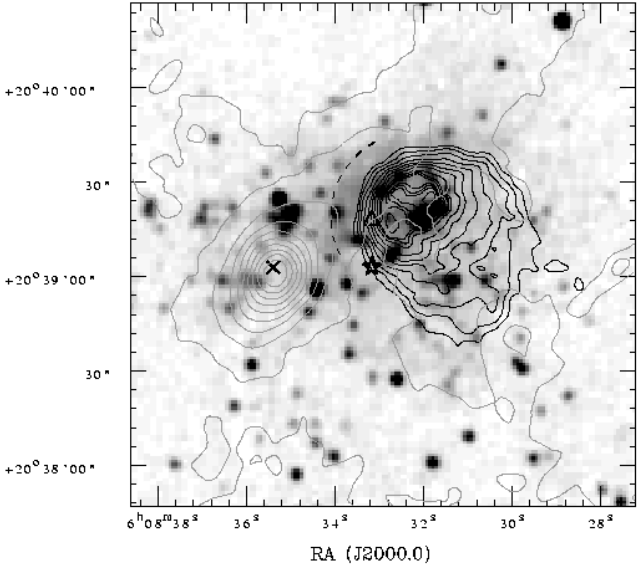}}
\caption{Contour maps of the emission from the ionized gas at 610
MHz (thick line) and dust emission at 850\,$\mu$m (thin line) are
overlaid on the 2MASS $K_{s}$ band image for the region around IRAS
06055+2039. The dashed arc represents the position of shocked
H$_{2}$. The peak positions of the
IRAS-HIRES 12\,$\mu$m (open triangle) and 100\,$\mu$m (open star) maps are
indicated. The cross marks the position of the methanol and water
masers.}
\label{all.fig}
\end{figure}

In Fig.\,\,\ref{all.fig}, we present the various components of the
region associated with IRAS 06055+2039. The plot displays the contour 
maps of the ionized gas and the emission from dust overlaid on the
2MASS $K_{s}$-band image. We show the contour plots of 610 MHz radio
emission, 850\,$\mu$m cold dust emission and mark the peak positions 
for the 100 and 12\,$\mu$m emission
from warm dust. The 850\,$\mu$m emission core lies to the S-E of the
ionized region with the warmer dust in between. The 12 and 
100\,$\mu$m peaks lie relatively closer to the radio peak. 
The ionized region is seen to be close to the edge
of the molecular cloud. Comparing with
Fig.\,\,\ref{h2_brg_rad.fig}, we note that the shocked
molecular gas lies in between the ionized region and the dense
molecular core. The central region of the infrared cluster is located
within the HII region which is at the edge of the molecular cloud. 

The ionized region and the dense molecular core as seen in the
radio and the sub-mm maps respectively are possibly at different
stages of evolution. The dense molecular core seen in our
sub-mm maps does not show any radio emission (down to level of the rms
noise which is 0.4 mJy/beam). The sub-mm peak is spatially offset from
the FIR peaks and there are no MSX or NIR counterparts seen. This
indicates a very early evolutionary stage for this dense and massive molecular core. It
is most probably a very early protocluster candidate and we are
sampling the initial collapse phase of the star forming core before
the formation of the UCHII region (Williams et al. 2004 and references
therein). This fact is further supported by the positions of the water and methanol 
masers which are seen to be coincident with the peak of the molecular core. 
The central peak of the CS map (Zinchenko et al. 1998)
and the peaks of the 450 and 850\,$\mu$m JCMT-SCUBA maps match with
the position of the masers.
More specifically, it is known that methanol maser sites are
generally radio quiet (as is the case here) and trace high mass star forming protoclusters
in very early evolutionary phases (Minier et al. 2005).  
On the other hand, the ionized region which is associated with IRAS
06055+2039, has FIR emission, free-free radio emission and has NIR and
MSX counterparts. This could probably indicate that the source is at a later evolutionary stage
in between an evolved protocluster and a young cluster. Here, massive
stars have started forming and a detectable HII region has been
created. The cluster is partially embedded in the parental cloud. The
sub-mm emission is weak here. 
A small subcluster is seen close to the edge of the sub-mm core spatially
coincident with the secondary peak mentioned in
Sect.\,\,\ref{ssnd.sect}. 
Thus, from this multiwavelength study of
the region associated with IRAS 06055+2039, we see the signatures of 
different evolutionary stages at different locations. 

\section{Summary}
\label{concl.sec}

The massive star forming region associated with IRAS 06055+2039 has
been studied in detail in the infrared, radio and sub-mm
wavelengths which lead to the following conclusions.
\begin{itemize}
\item[1.]
The morphological details of the environment around IRAS
06055+2039 show that we have probed different stages of evolution
of star formation present in this cloud complex.
\item[2.]
High sensitivity and high resolution radio continuum maps at 1280 and
610 MHz obtained from our observations using GMRT show interesting cometary morphology.
Apart from the diffuse emission, the 1280 MHz
continuum map also shows the presence of several discrete sources
which probably represent high density clumps.
The total integrated emission implies an exciting star of spectral
type B0 -- B0.5 which is consistent with the estimates derived
from NIR colours and IRAS PSC flux densities.
\item[3]
The dense molecular cold core has been probed with JCMT-SCUBA at 450
and 850\,$\mu$m. The sub-mm peak is spatially offset from the peak of
radio emission. The sub-mm emission is most probably
from a very early protocluster candidate and the radio emission is
from a region at a later evolutionary stage where the massive stars have formed a HII
region though the cluster is still partially embedded in the prenatal
cloud.
\item[4.]
In the NIR, the Br$\gamma$ emission correlates well
with the radio continuum emission.
The H$_{2}$ (1-0) S1 line of molecular hydrogen, which traces the
first shocked neutral layer beyond the ionization front is seen as an
arc towards the N-E of the central IRAS point source and envelopes
the ionized emission mapped at radio and NIR wavelengths. The shocked
molecular gas lies between the ionized region and the dense cloud
core.
\item[5.]
Using the 2MASS data, we derive a power law slope of
0.43$\pm$0.09 for the KLF of the NIR cluster associated with this star
forming complex. This is consistent with the values obtained 
for other young embedded clusters. We estimate an age of 2 -- 3 Myr for
the cluster. The physical structure of the cluster suggests
that it is not yet in complete dynamical equilibrium which is
consistent with the scenario of different evolutionary stages seen in
the complex.
\item[6.]
The spatial distribution of the emission from the UIBs as extracted from
the MSX images also displays cometary morphology. The total UIB
emission from MSX band A is estimated to be $\rm 2.85\times10^{-12} \rm Wm^{-2}$ 
and is $\la$ 39\% of the emission estimated from the IRAS LRS
spectrum.
The MSX mid-infrared colours of the central bright IRAS point source 
corresponds to a compact HII region.
\item[7.]
The spatial distribution of temperature and optical depth of
interstellar dust has been presented based on the mid- and far- infrared
data from the MSX and IRAS (HIRES) surveys. 
From the derived peak value of the $\tau_{100}$ distribution, we
estimate the warm dust mass
to be $\sim$ 6 $\rm M_{\odot}$.
From the sub-mm emission at 450 and 850\,$\mu$m, we estimate the total mass
of cloud to be $\sim$ 7000 -- 9000 $\rm M_{\odot}$. 

\end{itemize}

\subsection*{Acknowledgments}
We thank the anonymous referee for the comments and
suggestions which helped in improving the paper.
We thank the staff at the GMRT who have made the radio observations
possible. GMRT is run by the National Centre for Radio Astrophysics of
the Tata Institute of Fundamental Research. We thank IPAC, Caltech for
providing us the HIRES-processed IRAS products.
We thank Annie Robin for letting us use the Besan\c{c}on Galactic model. 
%-----------------------------------------------------------------------------

\end{document}